\documentclass[prb,twocolumn,superscriptaddress]{revtex4-1}
\usepackage{amsthm,amsmath,appendix,amsfonts,bm,caption,color,float,graphicx,subcaption,lipsum,verbatim,youngtab,lipsum}

\usepackage[colorlinks,bookmarks=true,citecolor=blue,linkcolor=blue,urlcolor=blue]{hyperref}
\theoremstyle{definition}

\newtheorem{theorem}{Theorem}[section]

\newtheorem{conjecture}{Conjecture}[theorem]
\newtheorem{question}{Question}[theorem]
\newcommand{\p}{\partial}
\newcommand{\Kl}{K\"{a}hler~}
\newcommand{\tcp}{\mathbb{CP}}
\newcommand{\cpone}{$\tcp^1$~}
\newcommand{\cptwo}{$\tcp^2$~}
\newcommand{\tcpone}{\tcp^1}
\newcommand{\tcptwo}{\tcp^2}
\newcommand{\tSU}{\text{SU}}
\newcommand{\tS}{\text{S}}
\newcommand{\tU}{\text{U}}
\newcommand{\ti}{\text{i}}
\newcommand{\ty}{\text{y}}

\begin{document}
\title{Fractional Quantum Hall States on \cptwo Space}

\author{Jie Wang}
\email{jiewang@flatironinstitute.org}
\affiliation{Center for Computational Quantum Physics, Flatiron Institute, 162 5th Avenue, New York, NY 10010, USA}
\author{Semyon Klevtsov}
\email{klevtsov@unistra.fr}
\affiliation{IRMA, Universit\'e de Strasbourg, UMR 7501, 7 rue Ren\'e Descartes, 67084 Strasbourg, France}
\author{Michael R. Douglas}
\email{mdouglas@scgp.stonybrook.edu}
\affiliation{Center of Mathematical Science and Applications, Harvard University, USA}
\affiliation{Department of physics, YITP and SCGP, Stony Brook University, USA}

\begin{abstract}
We study four-dimensional fractional quantum Hall states  on \cptwo geometry from microscopic approaches. While in 2d the standard Laughlin wave function, given by a power of Vandermonde determinant, admits a product representation in terms of the Jastrow factor, this is no longer true in higher dimensions. In 4d we can define two different types of Laughlin wavefunctions, the Determinant-Laughlin (Det-Laughlin) and Jastrow-Laughlin (Jas-Laughlin) states. We find that they are exactly annihilated by, respectively, two-particle and three-particle short ranged interacting Hamiltonians. We then mainly focus on the ground state, low energy excitations and the quasi-hole degeneracy of Det-Laughlin state. The quasi-hole degeneracy exhibits an anomalous counting, indicating the existence of multiple forms of quasi-hole wavefunctions. We argue that these are captured by the mathematical framework of the ``commutative algebra of N-points in the plane''. We also generalize the pseudopotential formalism to dimensions higher than two, by considering coherent state wavefunction of bound states. The microscopic wavefunctions and Hamiltonians studied in this work pave the way for systematic study of high dimensional topological phase of matter that is potentially realizable in cold atom and optical experiments.
\end{abstract}

\maketitle

\section{Introduction}
Searching for and understanding exotic phases of matter is a long standing goal of modern condensed matter physics. In three or lower spatial dimensions, many materials exhibiting topological or exotic properties have been synthesized. Higher dimensions, although realizable using optical and cold atom experimental techniques, are less well understood. It is both theoretically and experimentally interesting to ask about the possible many-body phenomena in four and higher dimensions \cite{HuZhang01,KARABALI02}.

The quantum Hall effect is the most studied phase of matter exhibiting topological properties \cite{RevModPhys.71.S298,RevModPhys.82.3045}. It was discovered forty years ago in electron gases confined in two dimensional semiconductor heterostructure in the presence of an ultra strong magnetic field \cite{RevModPhys.71.S298}. In magnetic fields, an electron gas reorganizes itself into completely dispersionless Landau levels, in which the complete quench of kinetic energy paves the way for purely interaction driven physics. Interactions give raise to exotic many-body states including the Laughlin phase \cite{laughlin}, which exhibits fractional Hall conductivity, and fractionalized particles or anyons \cite{RevModPhys.80.1083}. Theoretically Landau levels have been predicted to exist in even dimensions \cite{HuZhang01,KARABALI02,Karabali_2006,Karabali16,Karabali20,Douglas2010,4dqhedge_ergodic,4dqh_particlecp3}, and examples have been realized in optical and cold atom experiments \cite{review_syndim,4dqh_coldatom,Zilberberg:2018aa,Lohse:2018aa,4dqh_quasicrystal}.

Motivated by these studies, in this work we explore interacting effects in four and higher dimensions with uniform magnetic fields by initiating the study of the fractional quantum Hall (FQH) problem on $\tcp^n$ geometry. We start by microscopically describe the \cptwo geometry and the Landau levels. We then begin the discussion of many-body wavefunctions with a focus on Laughlin states and their quasi-hole descendants. In particular, we found there exist multiple types of Laughlin wavefunctions that are characterized by two-particle or three-particle clustering properties on the \cptwo space:
\begin{eqnarray}
\Psi^D_{\beta} &=& \left(\det_{1\leq i,j\leq N} f_{i,j}\right)^{\beta},\quad f_{i,j}\equiv \prod_{a=1}^3(z_i^a)^{p^a_j}\label{Det_Laughlin}\\
\Psi^J_{\gamma} &=& \left(\prod_{1\leq i<j<k\leq N}\epsilon_{abc}z^a_iz^b_jz^c_k\right)^{\gamma}.\label{Jastrow_Laughlin}
\end{eqnarray}

Here each particle's position on \cptwo is specified by homogeneous coordinates, {\it i.e.} a complex triplet $\bm z_i=(z^1,z^2,z^3)_i$ where subscript $i=1,...,N$ labels the particle. In Eqn.~(\ref{Det_Laughlin}), $\det f$ is a ``Slater'', or ``generalized Vandermonde'', determinant, determined by a set of $N$ triples of non-negative integers $\bm p_j=(p^1,p^2,p^3)_j$. Notations are specified in detail in Section.~\ref{section:cp2ll}.

The constructions of Eqn.~(\ref{Det_Laughlin}) and Eqn.~(\ref{Jastrow_Laughlin}) can be intuitively thought of as follows: while the Jastrow factor equals the Vandermonde determinant in two dimensions (for instance on $\tcp^1$), this is not true in higher dimensions. Viewing the \cpone Laughlin states as determinants or as Jastrow factors leads to distinct generalizations to $\tcp^2$. Thereby, we term Eqn.~(\ref{Det_Laughlin}) as Determinant-type Laughlin wavefunction (Det-Laughlin), and Eqn.~(\ref{Jastrow_Laughlin}) as Jastrow-type Laughlin wavefunction (Jas-Laughlin). We further show they are exact zero energy ground states for short ranged two-particle and three-particle repulsive interactions. In the end, we discuss how to classify generic transnational symmetric interactions by generalizing Haldane pseudopotentials \cite{haldanehierarchy,GreiterCP1} to high dimensions \cite{CHChen_PRL07,CHChen_AOP07,CHChen_PRB10}.

The quasi-hole excitations show interesting differences when compared with two dimensions. We found huge degeneracies in the quasi-hole space on \cptwo manifold, which might be useful for storing quantum information \cite{KITAEV20032}. Mathematically, the degeneracy of quasi-hole wavefunctions is interesting, and we will compare our results with relevant work of Haiman {\it et al.} \cite{Haiman}.

The many-body wavefunction and model Hamiltonian developed here serves as an explicitly solvable model of a higher dimensional topologically ordered system \cite{XGW_review,Janowitz:2013aa,PengYe2104}, less well known than its lower dimensional analogs \cite{Wen_TO_rigidstates,Wen_TO_edgeFQH,Wen_Int}, yet potentially realizable in cold atom and optical experiments \cite{review_syndim,4dqh_coldatom,Zilberberg:2018aa,Lohse:2018aa,4dqh_quasicrystal}. Further studying collective modes and numerically detecting non-point like excitations \cite{BERNEVIG2002185,4dqh_membrane} are just one of the interesting future directions.

The paper is organized as follows. We begin in Section.~\ref{section:cp2ll} with an introduction to \cptwo Landau levels. We also define coherent state representation \cite{haldanehierarchy,GreiterCP1,RMP_coherentstate} of single-particle and multi-particle bound-state wavefunctions in this section. In Section.~\ref{section:laughlin}, we define two types of Laughlin wavefunctions and discuss their parent Hamiltonians. In Section.~\ref{section:laughlinhole}, we numerically study the low energy excitations as well as the quasi-hole excitations of the Det-Laughlin state. In the subsequent section Section.~\ref{section:pp}, we discuss symmetric interaction and the higher dimensional generalization of Haldane-pseudopotentials \cite{haldanehierarchy,GreiterCP1,CHChen_PRL07,CHChen_AOP07,CHChen_PRB10}. Finally in Section.~\ref{section:math}, we discuss the connection between our studies to the commutative algebra in the mathematical literature. We list open questions and interesting future directions in Section.~\ref{section:discussion}.

\section{\cptwo Landau level\label{section:cp2ll}}
The two-sphere $\tS^2\simeq\tcpone$ geometry is one of the most useful geometries for studying two dimensional quantum Hall physics \cite{haldanehierarchy,GreiterCP1}. The discussion of higher dimensional quantum Hall physics was initiated by S.C. Zhang and J. Hu in Ref.~(\onlinecite{HuZhang01}), who gave single particle Landau level wavefunctions on the four-sphere $\tS^4$. Soon after, D. Karabali  and  V. Nair, generalized the $S^2$ Landau levels to the $2n$ dimensional complex projective spaces $\tcp^n$ in Ref.~(\onlinecite{KARABALI02}) and the generic case of a compact \Kl manifold with a uniform magnetic field was treated in Ref.~(\onlinecite{Douglas2010}). In this section, we first review Landau levels on \cptwo space, introduce a diagrammatic representation, and discuss coherent states which turn out to be useful for discussing interactions.

\subsection{\cptwo geometry and lowest Landau level states}
The \cptwo manifold is conveniently parameterized by three complex variables, satisfying a real constraint and identifying points which are equivalent up to a $\tU(1)$ phase:
\begin{eqnarray}
&&\bm z = (z^1,z^2,z^3) = \left(u, v, w\right),\label{triplet}\\
&&|u|^2+|v|^2+|w|^2 = 1, \quad \bm z \sim \bm z e^{i\theta},\nonumber
\end{eqnarray}
where a subscript $i=1,...,N$ will be added to label particles when discussing many-particle physics. Throughout this work, we will use superscript $a=1,2,3$ to denote the inner index of $\tSU(3)$ quantum number, and we will use $(z^1,z^2,z^3)$ and $(u,v,w)$ interchangeably.

\cptwo has an $\tSU(3)$ symmetry under which these coordinates transform in the fundamental representation.  There is a unique metric (up to the overall scale) which respects this symmetry, the Fubini-Study metric \cite{Douglas2010}. The analogous expression with an $(n+1)$ component vector parameterizes $\tcp^n$, with $\tSU(n+1)$ symmetry.

\cptwo can also be obtained as the homogeneous space $\tcptwo = \tSU(3)/\left[\tSU(2)\times \tU(1)\right]$. This construction defines a natural background $\tSU(2)\times \tU(1)$ gauge field with minimal magnetic charge under the Dirac quantization condition and which respects the $\tSU(3)$ symmetry.  The $\tU(1)$ part of this field has magnetic field strength given by a two-form $\mathcal{F}_{ab}$, which stands in a simple mathematical relationship to the Fubini-Study metric: it is the \Kl form for this metric. It is this relation which is responsible for the simple form of the lowest Landau level (LLL) wavefunctions, and comparable results could be obtained for any \Kl manifold, as derived in Ref.~(\onlinecite{KARABALI02}).

In this work we will consider particles with charge $e=S$ under the $\tU(1)$ Abelian magnetic field, and with zero $\tSU(2)$ charge. Equivalently, we can think of charge $e=1$ particles under the influence of a magnetic field with flux $S$.  

Following Ref.~(\onlinecite{KARABALI02}), the LLL wavefunctions are holomorphic functions of $\bm z=\left(u,v,w\right)$, with no dependence on the complex conjugation $\bar{\bm z}$. A complete orthonormal basis for such functions is
\begin{equation}
\psi_{S,\bm p} =  u^{p^1}v^{p^2}w^{p^3},\label{singleparticlewf}
\end{equation}
where $\bm p=(p^1,p^2,p^3)$ is an vector of non-negative integers
and we define $p^1+p^2+p^3=S$.  The subset with fixed $S$ are the
wavefunctions with that $\tU(1)$ charge, of total number $D=(S+1)(S+2)/2$.

Instead of $\bm p$, we sometimes use the labels $(i,y)$, {\it i.e.} ``isospin'' and ``hypercharge'', defined by:
\begin{eqnarray}\nonumber
\ti &=& p^1-p^2,\quad \ty = p^1+p^2-2p^3.
\end{eqnarray}

For an $N-$particle many-body state, we use capital $I = \sum_{i=1}^N\ti_i$, $Y = \sum_{i=1}^N\ty_i$ to label the total ``isospin'' and ``hypercharge''. The total $\sum_{i=1}^{N}p^{1,2}_i = (2NS+Y\pm 3I)/6$ are integers. Therefore,
\begin{equation}
    2NS + Y \pm 3I \in 6\mathbb{Z}.\label{constrain}
\end{equation}

\subsection{Diagrammatic Representation and Bivariate Vandermonde Determinant}
Here we introduce a diagrammatic representation for \cptwo states. We represent the quantum numbers $p^1$ and $p^2$ by horizontal and vertical axes respectively.   Since $p^{1,2,3}\in[0,S]$, each \cptwo state is represented as a box in a triangle of base and height $S+1$. 
We then represent a particular $N$ particle state by distributing
$N$ dots among the boxes, one per particle.  If the particles are
fermions, the Pauli principle is enforced by allowing at most one
dot per box.

For example, the one particle state with $S=2$ and $\bm p=(0,1,1)$ is
depicted by:
\begin{equation}\label{diag1}
\young(\hfil,\cdot\hfil,\hfil\hfil\hfil)
\end{equation}

As another example, consider $N=2$ particles and the $S=2$,
$I=1$, $Y=1$ subspace.  There are two basis states, depicted by:
\begin{equation}\label{example_N3S2}
\young(\hfil,\cdot\hfil,\hfil\hfil\cdot)\quad\young(\hfil,\hfil\cdot,\hfil\cdot\hfil)
\end{equation}

The fermionic wavefunction corresponding to a diagram can be written as a \emph{bivariate determinant}, a two variable generalization of the usual Vandermonde determinant. Denoting the set of filled boxes as $\{\bm p^1,\ldots,\bm p^N\}$, the corresponding $N$ particle wavefunction is
\begin{eqnarray}
\Delta_{\{\bm p\}} &=& \det_{i,j}\left(\bm z_i^{\bm p^j}\right),\nonumber\\
&=& \det \begin{bmatrix} u_1^{p^1_1}v_1^{p^2_1}w_1^{p^3_1} & ... & u_1^{p^1_N}v_1^{p^2_N}w_1^{p^3_N} \\ \vdots & & \vdots \\ u_N^{p^1_1}v_N^{p^2_1}w_N^{p^3_1} & ... & u_N^{p^1_N}v_N^{p^2_N}w_N^{p^3_N} \\ \end{bmatrix}.\label{eq:bivandet}
\end{eqnarray}

Eqn.~(\ref{eq:bivandet}) is indeed ``bivariate'' (rather than tri-variate) because $u,v,w$ are constrained by $|u|^2+|v|^2+|w|^2=1$. To see this more explicitly, one can turn ``homogeneous coordinates'' $\bm z=(u,v,w)$ into ``projective coordinates'':
\begin{equation}
    \tilde u\equiv u/w,\quad \tilde v\equiv v/w,\label{proj_coord}
\end{equation}
and Eqn.~(\ref{eq:bivandet}) can be rewritten as,
\begin{eqnarray} \label{eq:bivandet2}
\Delta_{\{\bm p\}} &=& \left(\prod_{i=1}^Nw_i^S\right) \det \begin{bmatrix} \tilde u_1^{p^1_1}{\tilde v}_1^{p^2_1} & ... & \tilde u_1^{p^1_N}{\tilde v}_1^{p^2_N} \\ \vdots & & \vdots \\ \tilde u_N^{p^1_1}{\tilde v}_N^{p^2_1} & ... & \tilde u_N^{p^1_N}{\tilde v}_N^{p^2_N} \\ \end{bmatrix}.
\end{eqnarray}

We sometimes use captical $Z\equiv(\tilde u, \tilde v)$ to denote the projective coordinate.

\subsection{Coherent States}
Instead of discrete integer-valued quantum numbers $(p_1,p_2,p_3)$ or $(i,y)$, coherent states provide an overcomplete basis with continuous parameters \cite{RMP_coherentstate}. They are useful for considering multi-particle interactions \cite{haldanehierarchy,GreiterCP1}.

In the following discussion we use the notation $(z^1,z^2,z^3)$ in place of $(u,v,w)$. Bold face $\bm z$ is still used to represent a vector, while $\bar{\bm z}$ represents the vector obtained by complex conjugating each component. The dot product represents the standard Euclidean inner product, for example $\bar{\bm\alpha}\cdot \bm z = \bar\alpha^1z^1+\bar\alpha^2z^2+\bar\alpha^3z^3$.

\subsubsection{One-particle coherent state}
The single-particle coherent state wavefunction is parameterized by a \cptwo point $\bm\alpha=(\alpha^1,\alpha^2,\alpha^3)$,
\begin{equation}\label{singleparticlecoherent}
\psi^{(1)}_{S,\bm\alpha}(\bm z) = \left(\bar{\bm\alpha}\cdot \bm z\right)^{S},
\end{equation}
where just as in Eqn.~(\ref{triplet}), $\bm\alpha$ is normalized to one.
The point corresponding to an equivalence class of $\bm\alpha$
under $\tU(1)$ phase rotation is the point on which the coherent state is maximized, and in the limit $S\rightarrow\infty$ the state is localized at this point.

The $\tSU(3)$ symmetry acts on the coherent state by moving its center $\bm\alpha$ while keeping its shape invariant. Varying the polarization vector $\bm\alpha$ continuously yields an Abelian Berry phase corresponding to the $\tU(1)$ magnetic flux described earlier.

\subsubsection{Three-particle coherent state}
We next seek the coherent representations of bound states.
A bound state has a center of mass $\bm\alpha$ which transforms
as a fundamental of the $\tSU(3)$ global symmetry, and internal
variables which can also transform.  Given $\bm\alpha$ there is
a ``little group'' $\tSU(2)\times \tU(1)\subset \tSU(3)$ which preserves
$\bm\alpha$, which can be used to define two internal quantum
numbers.  One is the spin $J_1$ under this $\tU(1)$, which acts as \cite{Luis_2008,Luis05PRA1,Luis05PRA2},
\begin{equation}
\left(\sum_{i=1}^8\lambda_i(\bm\alpha)\cdot\hat{\bm\Lambda}_i\right)\psi\equiv J_1 \psi,\nonumber
\end{equation}
where $\hat{\bm\Lambda}_{i}$ are the Gell-Mann matrices, and $\lambda_i = \sum_{a,b=1}^3\bar\alpha^a\left(\hat{\bm\Lambda}_i\right)_{ab}\alpha^b$. The second is the $\tSU(2)$ Casimir $J_2$:
\begin{equation}
\left(\sum_{i=1}^8\hat{\bm\Lambda}_i\right)^2\psi\equiv J_2 \psi.\nonumber
\end{equation}

To construct bound states with definite values of these quantum numbers,
we use invariant tensors.  Now the only invariant
tensor which couples fundamental representation of $\tSU(3)$ is
the three-index tensor $\epsilon^{abc}$.  Because of this,
on \cptwo the three particle bound state is simpler than
the two particle bound state.

We define the ``three particle coherent state'' to be:
\begin{eqnarray}\label{threeparticlecoherent}
\psi^{(3)}_{S,J,\bm\alpha}(\bm z_{1},\bm z_{2},\bm z_{3}) &\equiv& \left(\epsilon_{abc}z^a_1z^b_2z^c_3\right)^{S-J} \prod_{i=1}^{3}\left(\bar{\bm\alpha}\cdot \bm z_i\right)^{J}. \nonumber
\end{eqnarray}

As its relative part $\epsilon_{abc}z^a_1z^b_2z^c_3$ is invariant under the little group, both of the internal quantum numbers are determined by $J$: 
\begin{equation} \label{eq:internalqs}
J_1=J; \qquad J_2=\frac{1}{3}J(J+3).
\end{equation}

\subsubsection{Two-particle coherent state}
This expression can be adapted to describe two particles on $\tcp^2$
by replacing the position $\bm z_3$ of the third particle with a
constant vector $\bm\beta$.  In some sense it describes the precession of the relative coordinate.

We define the ``two particle coherent state'' as:
\begin{eqnarray}\label{2pcoherentwf}
\!\!\!\psi^{(2)}_{S,J,\bm\alpha,\bm\beta}(\bm z_{1},\bm z_{2}) &\equiv& \left(\epsilon_{abc}z^a_1 z^b_2\beta^c\right)^{S-J} \prod_{i=1}^{2}\left(\bar{\bm\alpha}\cdot \bm z_i\right)^{J}.
\end{eqnarray}

It is not invariant under the little group. An invariant state can be made by averaging over the relative vector $\bm\beta$, and it has the same internal quantum numbers Eq.~(\ref{eq:internalqs}).

We will use this decomposition to develop a pseudopotential formalism 
and discuss generic symmetric interactions
in Section.~\ref{section:pp}.

\section{Laughlin Wavefunctions \label{section:laughlin}}
We proceed to discuss interacting physics. 
An $N-$particle Laughlin wavefunction will be defined to
be a totally antisymmetric (for $\beta$ odd) or symmetric (for $\beta$ even) state on the LLL, which vanishes to order $\beta$ when any pair of particles coincides.

On $\tcpone$, the $N-$particle Laughlin wavefunction is defined with $S=\beta(N-1)$ total degree. Its wavefunction reads:
\begin{eqnarray}
    \Psi_{\beta} &=& \prod_{i<j}^N\left(u_iv_j-u_jv_i\right)^\beta,\label{cponelaughlin_jast}\\
    &=& \det \begin{bmatrix} u_1^0v_1^{N-1} & ... & u_1^{N-1}v_1^{0} \\  \vdots & & \vdots \\ u_N^0v_N^{N-1} & ... & u_N^{N-1}v_N^{0} \\ \end{bmatrix}^{\beta}.\label{cponelaughlin_vand}
\end{eqnarray}

Eqn.~(\ref{cponelaughlin_jast}) and Eqn.~(\ref{cponelaughlin_vand}) are respectively written in a form of the Jastrow factor, and the Vandermonde determinant. The equivalence between Jastrow factor and Vandermonde determinant, however, is no longer true in higher dimensions. As will be seen in this section, this leads to two types of Laughlin wavefunctions on $\tcp^{n>1}$.

Nor is it {\it a priori} clear that either type of wavefunction completely exhausts the possible Laughlin wavefunctions. In Section.~\ref{section:math} we will show that the determinantal wavefunctions do cover all of the Laughlin wavefunctions.

\subsection{Type-I: Determinant-Laughlin Wavefunction}
We first discuss generalizing Laughlin wavefunction to four dimensions by using bivariate Vandermonde determinants. For degree $S$, the filled Landau level has in total $N=(S+1)(S+2)/2$ particles. We denote the filled Landau level wavefunction by $\Delta_{\text{FLL},S}$. For instance, the filled Landau level wavefunction for $S=2$ is represented as the following diagram, which involves $N=6$ particles:
\begin{equation}
\Delta_{\text{FLL},S2} = \Psi_{N6S2} = \young(\cdot,\cdot\cdot,\cdot\cdot\cdot)\label{S2filledLL}
\end{equation}

We define the Determinant type Laughlin wavefunction at filling fraction $\nu=\beta^{-1}$ as:
\begin{equation}
    \Psi_{\beta}^D = \left(\Delta_{\text{FLL},S/\beta}\right)^{\beta},\label{detlaughlin}
\end{equation}
which occurs when,
\begin{equation}
    N = \left(S/\beta+1\right)\left(S/\beta+2\right)/2.\label{NS_det}
\end{equation}

The Det-Laughlin wavefunctions vanishes in power of $\beta$ when any two particles approach each other. Following similar arguments of Trugman and Kivelson \cite{Trugman_Kivelson}, it is the exact zero energy ground state for short-ranged repulsive interactions. The concrete form of interacting Hamiltonian is given as follows in Eqn.~(\ref{interactH}), which is generalized from the two-dimensional form proposed by Wen {\it et al.} \cite{Wen_Int}:
\begin{equation}
    H_{\beta}^D = -\sum_{i<j}\sum_{l=0}^{\beta-1} V_{l} \left(\partial^{\dag}\right)^l\delta(Z_i, Z_j)\left(\partial\right)^l,\label{interactH}
\end{equation}
where $V_l$ are arbitrary non-negative potentials and the conjugation is defined with respect to an appropriate $L^2$ structure on many-body states. The $Z = (\tilde u,\tilde v)$ are the projective coordinates introduced in Eqn.~(\ref{proj_coord}).

\subsection{Type-II: Jastrow-Laughlin Wavefunction}
Alternatively, the \cpone Laughlin wavefunction can be regarded as a Jastrow factor as shown in Eqn.~(\ref{cponelaughlin_jast}). Generalizing the singlet $(u_iv_j-u_jv_i)$ to \cptwo requires three particles. Consider particle $1,2,3$, the singlet wavefunction is:
\begin{equation}\nonumber
    u_{123} \equiv \epsilon^{ijk}u_iv_jw_k;\quad i,j,k=1,2,3.
\end{equation}

We define the Jastrow type Laughlin wavefunction as:
\begin{equation}
\Psi_{\gamma}^J = \left(\prod_{i< j< k}u_{ijk}\right)^\gamma,\label{jaslaughlin}
\end{equation}
which occurs when,
\begin{equation}
S/\gamma = (N-1)(N-2)/2.\label{NS_jas}
\end{equation}

To be concrete, the $N=3,4,5$ particle Jas-Laughlin wavefunctions are:
\begin{eqnarray}
S/\gamma=1,\quad\Psi_{\gamma}^{J}(\bm u_{1,...,3}) &=& \left(u_{123}\right)^{\gamma},\label{JasLauWf}\\
S/\gamma=3,\quad\Psi_{\gamma}^{J}(\bm u_{1,...,4}) &=& \left(u_{123}~u_{124}~u_{134}~u_{234}\right)^{\gamma},\nonumber\\
S/\gamma=6,\quad\Psi_{\gamma}^{J}(\bm u_{1,...,5}) &=& \bigl(u_{123}~u_{124}~u_{125}~u_{134}~u_{135}~u_{145}\nonumber\\
&&~u_{234}~u_{235}~u_{245}~u_{345}\bigl)^{\gamma}.\nonumber
\end{eqnarray}

Note that we used $\gamma$ to parameter the wavefunction Eqn.~(\ref{jaslaughlin}). The two-particle vanishing power of Jas-Laughlin wavefunction, defined as the vanishing power when any pair of particles approach, is,
\begin{equation}
    \beta = \gamma(N-2).\label{vanishingpowerJastrow}
\end{equation}

Therefore, for both Det-Laughlin and Jas-Laughlin, $\beta$ determines the statistics: the wavefunction is fermionic if $\beta$ is odd, and bosonic if $\beta$ is even.

The vanishing power $\beta$ for Jas-Laughlin is tricky, as it dependents on the particle number $N$, therefore for fixed $\gamma$ there is no thermodynamic definition for the Jas-Laughlin based on $\beta$. The $\gamma$ instead defines the three-particle clustering property of Jas-Laughlin, and this phase is thereby characterized by three, rather than two, particle properties.

Throughout this work, when saying ``Laughlin'', we implicitly refers to the ``Det-Laughlin'', {\it i.e.} we will use the terminology ``Det-Laughlin'' and ``Laughlin'' interchangeably. For the Jastrow type Laughlin states, we will explicitly term them as ``Jas-Laughlin''.



\subsection{Comparing two Types of Laughlin Wavefunctions}
\begin{figure}[ht]
	\centering
	\includegraphics[width=1.0\linewidth]{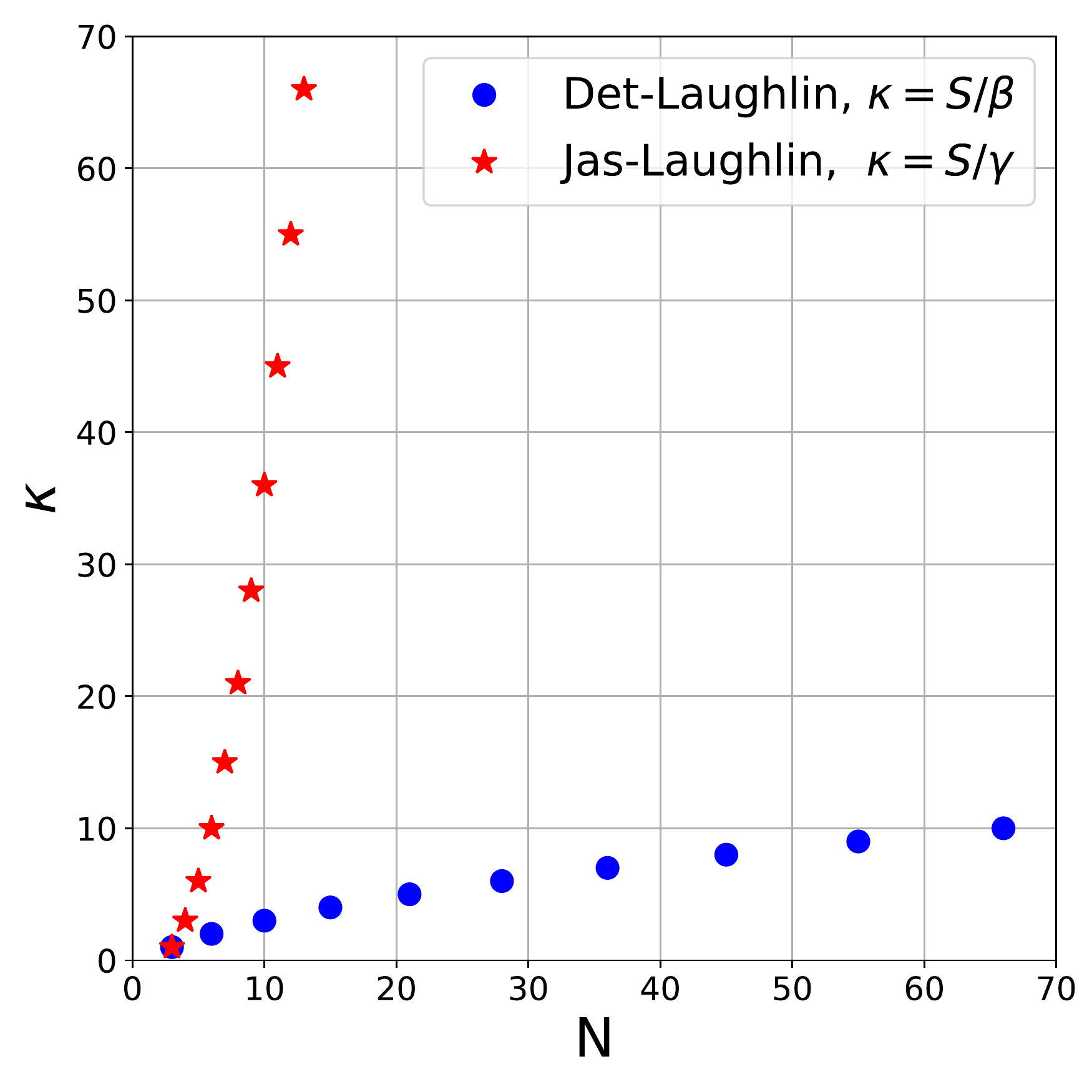}
	\caption{Illustration of Table.~\ref{table_laughlin}, from which we see directly Det-Laughlin generally is much denser than Jas-Laughlin.}
	\label{NSrelation}	
\end{figure}

We end this section by tabulating the particle number $N$ and the associated total degree that the two Laughlin wavefunctions can occur in Table.~\ref{table_laughlin}. Note that for $N=3$ particles, the two types of Laughlin wavefunction coincide. Generally speaking, $\Psi^D$ is denser than $\Psi^J$ as seen from FIG.~\ref{NSrelation}.

\begin{table}[h]
\centering
\begin{tabular}{p{1.1cm}|p{1.1cm}|p{1.4cm}|p{1.1cm}|p{1.1cm}|p{1.4cm}}
 \hline\hline
 \multicolumn{3}{ c |}{Det-Laughlin $\Psi^D_{\beta}$} & \multicolumn{3}{| c }{Jas-Laughlin $\Psi^J_{\gamma}$} \\
 \hline
 \hfil $N$ & \hfil $S/\beta$ & \hfil statistics & \hfil $N$ & \hfil $S/\gamma$ & \hfil statistics \\
 \hline
 \hfil 3  & \hfil 1 & \hfil $\beta\mod2$ & \hfil 3 & \hfil 1 & \hfil $\gamma\mod2$ \\
 \hfil 6  & \hfil 2 & \hfil $\beta\mod2$ & \hfil 4 & \hfil 3 & \hfil $0$ \\
 \hfil 10 & \hfil 3 & \hfil $\beta\mod2$ & \hfil 5 & \hfil 6 & \hfil $\gamma\mod2$ \\
 \hfil $...$ & \hfil $...$ & \hfil $...$ & \hfil $...$ \\
 \hline\hline
\end{tabular}
\caption{The particle number $N$ and the associated degree $S$ for the Det-Laughlin $\Psi^D_{\beta}$ and the Jas-Laughlin $\Psi^J_{\gamma}$, following Eqn.~(\ref{NS_det}) and Eqn.~(\ref{NS_jas}). ``Statistics'' is $0$ ($1$) means that the Laughlin state is realized by interacting bosons (fermions). For $N=3$ particles, Det-Laughlin is equivalent as the Jas-Laughlin. The $N=4$ Jas-Laughlin is the simplest nontrivial example which are zero energy ground state of $\sum_{i<j}\delta(Z_i-Z_j)$ interactions but cannot be written as a product of determinants.}\label{table_laughlin}
\end{table}

The Det-Laughlin wavefunction has been proved to be the unique zero energy ground state for two-particle short ranged interaction \cite{CHChen_PRL07}. We verify this result numerically in the next section. For this reason, we term the $(N,S/\beta)$ of the left column as the ``commensurate parameter'' for Det-Laughlin wavefunction. Similarly, the $(N,S/\gamma)$ of the right column can be termed as the commensurate parameter for Jas-Laughlin wavefunctions. Note that Jastrow wavefunctions need to be stabilized by three-particle rather than two-particle interactions.

In the next section, we numerically study the parent Hamiltonian of the Det-Laughlin, with a focus on its ground state, low-lying excitations and quasi-hole degeneracies.

\section{Numerical Studies \label{section:laughlinhole}}
\subsection{Numerical Diagonalization}
Laughlin wavefunctions are characterized by their clustering behavior: on \cpone the Laughlin wavefunction $\Psi_{\beta}$ vanishes in power of $\beta$ when any two particles coincide. This yields the consequence that Laughlin wavefunctions are exact zero energy ground states for any repulsive interactions as $\sum_{l<\beta}-V_l\partial^{\dag l}\delta(Z_i,Z_j)\partial^l$ where $V_l>0$.

In the following sections, we numerically study the Laughlin state and their quasi-hole descendent for repulsive two-body interaction at various $(N,S)$. For fermions, we use,
\begin{equation}
H=-\sum_{i<j}\partial^{\dag}\delta(Z_i,Z_j)\partial,\label{deltaprimeint}
\end{equation}
and for bosons, we use
\begin{equation}
H=\sum_{i<j}\delta(Z_i,Z_j).\label{deltaprimeintboson}
\end{equation}


As reviewed in the first section, orthonormal non-interacting single particle states in this Hilbert space are labeled by Eqn.~(\ref{singleparticlewf}). We first derive the two-body interaction element,
\begin{equation}
    V_{\bm q\bm q';\bm p\bm p'} \equiv \langle \bm q\bm q'|H|\bm p\bm p'\rangle,\label{matrixelements}
\end{equation}
where $|\bm p\bm p'\rangle$ is an anti-symmetrized two-particle wavefunction. With the matrix elements Eqn.~(\ref{matrixelements}), the second quantized Hamiltonian reads:
\begin{equation}
    H = \sum_{\bm p_1\bm p_2;\bm p_3\bm p_4}V_{\bm q\bm q';\bm p\bm p'}\delta_{\bm p+\bm q=\bm p'+\bm q'}c^{\dag}_{\bm q}c^{\dag}_{\bm q'}c_{\bm p}c_{\bm p'},\label{secondquantizedH}
\end{equation}
where $c^{\dag}_{\bm p}$ creates a single-particle wavefunction $\psi_{S,\bm p}$ as seen in Eqn.~(\ref{singleparticlewf}). The $\delta$ function above stems from the $\tSU(3)$ quantum number conservation since the interaction is $\tSU(3)$ symmetric (corresponding to transnational invariant in the thermodynamic limit). Diagonalizing the second quantized Hamiltonian Eqn.~(\ref{secondquantizedH}) gives many-body wavefunctions and energies. The matrix elements are straightforwardly derived using the single particle wavefunctions.

For all $(N,S)$ listed in the left-panel of Table.~\ref{table_laughlin} for Vandermonde Laughlin $\Psi_{\beta}^D$, numerically we found single degenerate many-body zero mode for interaction Eqn.~(\ref{deltaprimeint}). They correspond to the \emph{fermionic Det-Laughlin wavefunctions} at $\beta=3$. They are fermionic because the vanishing power $\beta$ is odd.

We also found a single zero mode at $N=4, S=3$ for interaction Eqn.~(\ref{deltaprimeintboson}). This corresponds to the four-particle \emph{bosonic Jas-Laughlin wavefunction} of $\gamma=1$ listed in the second line of Eqn.~(\ref{JasLauWf}). It is bosonic because its vanishing power $\beta$, according to Eqn.~(\ref{vanishingpowerJastrow}), is even. Interestingly, this wavefunction is the simplest wavefunction which exhibits the anomalous counting of \cptwo which we will discuss more in Section.~\ref{section:math}: this wavefunction \emph{cannot} be represented by a product of Vandermonde determinant, but rather is a linear combination of determinants where the high order vanishing powers has cancellation.

At $(N,S)=(5,6)$, we observed multiple zero modes for Eqn.~(\ref{deltaprimeint}). One of them is the $N=5, \gamma=1$ Jastrow wavefunction, {\it i.e.} the last line of Eqn.~(\ref{JasLauWf}). Besides, the zero mode space also include quasi-hole descendants of $(N,S)=(6,6)$ Det-Laughlin as $(N,S)=(5,6)$ can be obtained from $(N,S)=(6,6)$ by removing one particle.

\subsection{Low energy excitation of Det-Laughlin}
\begin{figure}[ht]
     \centering
     \includegraphics[width=1.0\linewidth]{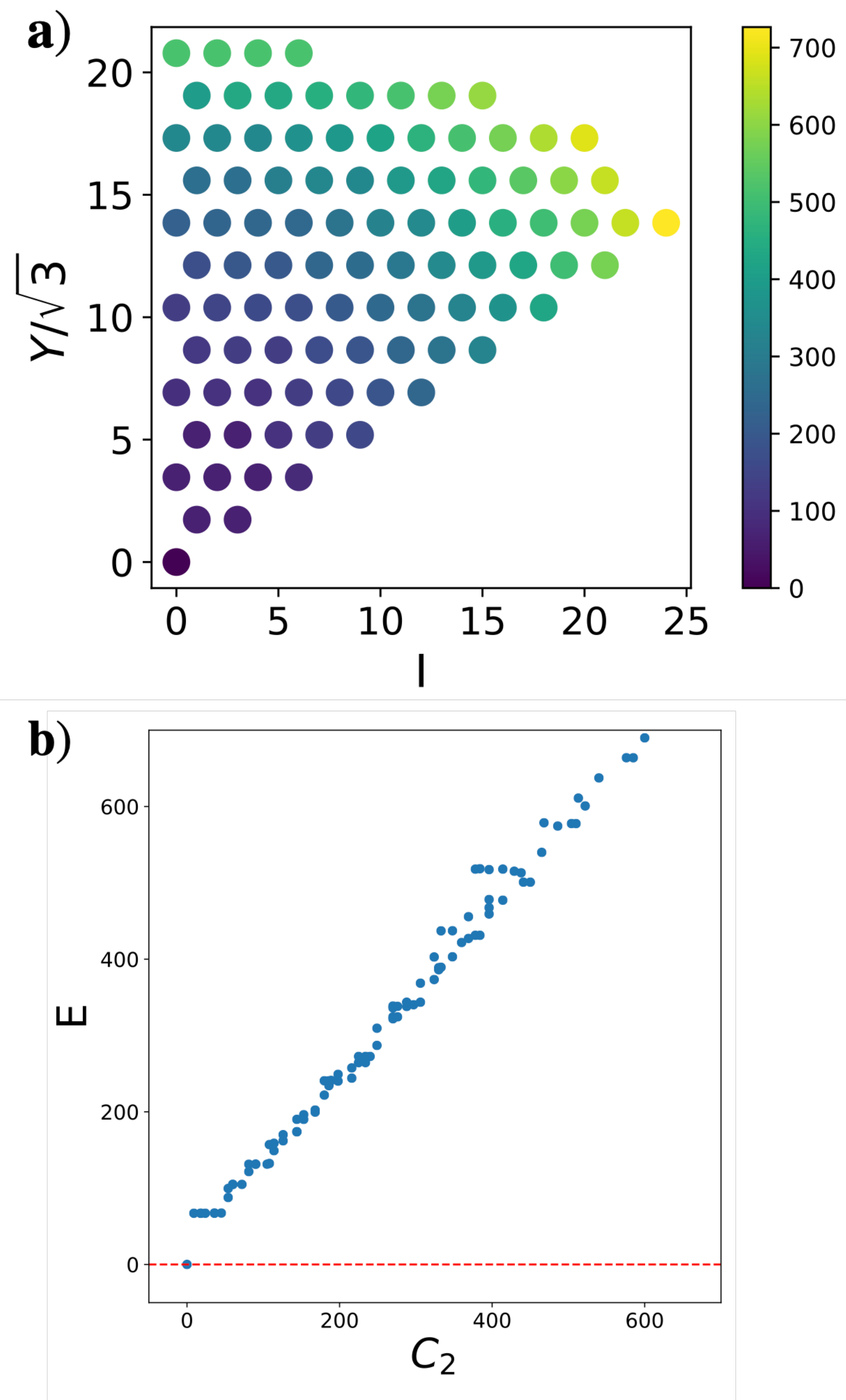}
     \caption{Ground state and low-energy excitations in $(N,S)=(6,6)$ sector. (a) shows the lowest energy in each quantum number sector. (b) plots the same data as (a) but in terms of the quadratic Casimir $C_2$ of $\tSU(3)$. In (b), a unique zero energy ground state at $I=Y=0$ is clearly seen. Moreover, (b) indicates the low energy excitations are well approximated by a quadratic dispersion form.}\label{N6S6numerics}
\end{figure}

In this section, we focus on the ground state and low energy excitation at the commensurate filling fraction for $N=6$ particle Det-Laughlin at $S=6$. As shown in FIG.~\ref{N6S6numerics}, the numerically observed unique zero energy ground state and a finite energy excitation may indicate the Det-Laughlin is an in-compressible state. Careful numerical studies about the finite size scaling of the gap are required in the future.

Since the interaction Eqn.~(\ref{deltaprimeint}) is $\tSU(3)$ invariant, it will take the same value on every state in an irreducible $\tSU(3)$ representation. This value can be written as a function of two Casimir invariants, the quadratic Casimir $C_2$ and the cubic Casimir $C_3$. For a given irreducible representation, there is a unique highest weight state (largest $I$ and $Y$), and one can write the Casimirs as functions of its quantum numbers. Particularly,
\begin{equation}
C_2 = 3I^2+Y^2+6(I+Y).
\end{equation}

As shown in FIG.~\ref{N6S6numerics}, the lowest energy in each quantum number sector $(I,Y)$ displays a good linearity, if plotting the quantum number in terms of the quadratic Casimir $C_2$. This indicates a quadratic low energy dispersion which may be captured by a high dimensional generalization of the magneto-roton theory \cite{gmpl,gmpb}, which we leave for future studies.



\subsection{Quasi-hole degeneracies}
We have numerically studied the energy spectrum of short-ranged interaction Eqn.~(\ref{deltaprimeint}) in the Hilbert space $(N,S)$ listed in the Table.~\ref{table_laughlin}. We found the Det-Laughlin are the unique ground states. While Jas-Laughlin are not generally the unique $E=0$ states for two-particle interactions, they are for three-particle interactions as we will discuss later.

In this section, we introduce one extra flux quanta from the commensurate parameter of Det-Laughlin, and study the ground state degeneracies of two-particle interaction Eqn.~(\ref{deltaprimeint}). The ground state degeneracies correspond to the dimension of quasi-hole wavefunctions. We found anomalous counting in the quasi-hole degeneracy, which has close connection to the mathematical subject discussed in Section.~\ref{section:math}.

\subsubsection{N3S4}
We start with the simplest quasi-hole state at $(N,S)=(3,4)$, which descends from the simplest Det-Laughlin wavefunction at $(N,S)=(3,3)$. In Table.~\ref{N3S4}, we list the dimension of the zero-modes in each quantum number $(I,Y)$ sector which counts the dimension of quasi-hole wavefunctions.

\begin{table}[h]
\centering
\begin{tabular}{p{1.5cm}|p{1.4cm}|p{1.4cm}|p{1.4cm}|p{1.4cm}}
\hline\hline
 & \hfil I=0 & \hfil I=$\pm$1 & \hfil I=$\pm$2 & \hfil I=$\pm$3 \\
 \hline
 \hfil Y=~~0 & \hfil 2 & & \hfil 2 & \\
 \hline
 \hfil Y=$\pm$3 & & \hfil 2 & & \hfil 1 \\
 \hline
 \hfil Y=$\pm$6  & \hfil 1 & & \hfil 0 & \\
 \hline\hline
\end{tabular}
\caption{Zero mode space dimension for Eqn.~(\ref{deltaprimeint}) interaction in $(N, S)=(3, 4)$ sector. All zero modes can be written as product of three determinants of degree $S=1+1+2$. Empty grids are invalid quantum numbers as they are constrained by Eqn.~(\ref{constrain}).}\label{N3S4}
\end{table}

In fact, in this case, quasi-hole zero modes are all given by the wavefunction which is written as product of three determinants of degree $S=1+1+2$:
\begin{eqnarray}
\Psi_{N3S4} = \left(\Delta_{\text{FLL},S1}\right)^2\times\Psi_{N3S2},\label{HoleWf_N3S4}\\
\nonumber
\end{eqnarray}
where $\Delta_{\text{FLL},S}$ is the fully filled \cptwo Landau level wavefunction defined above Eqn.~(\ref{S2filledLL}). The $\Psi_{N3S2}$ has three particles filled in space of degree $S=3$, and it is diagrammatically represented as shown in Eqn.~(\ref{example_N3S2}). From Eqn.~(\ref{example_N3S2}), we also see why the dimension of $(I,Y)=(0,0)$ sector in Table.~\ref{N3S4} is two. Degeneracies in other quantum number sectors can be worked out diagrammatically straightforwardly following the same spirit. This case does not show any anomalous properties as the dimensions of zero modes are all expected.

\subsubsection{N6S7}
We next study the zero mode dimension in the $(N,S)=(6,7)$ sector, which descends from the $(N,S)=(6,6)$ Det-Laughlin by adding one extra flux quanta. Analogies to Eqn.~(\ref{HoleWf_N3S4}), we first write down the quasi-hole wavefunctions as product of determinants of degree $S=2+2+3$:
\begin{eqnarray}
\Psi_{N6S7} = \left(\Delta_{\text{FLL},S2}\right)^2\times\Psi_{N6S3},\label{HoleWf_N6S7}\\
\nonumber
\end{eqnarray}
where $\Psi_{N6S3}$ represents diagrams of $N=6$ dots filled in degree $S=3$.

However, as we see in Table.~\ref{N6S7}, this type of quasi-hole wavefunctions is \emph{not} enough to explain the degeneracies observed numerically: in each $(I,Y)$ sector, we label the dimension of quasi-hole wavefunction of type Eqn.~(\ref{HoleWf_N6S7}) as the number outside the bracket, and we list in the bracket the additional zero modes dimension observed numerically. The total zero mode space dimension seen numerically is $266$, which has in total $56$ more zero modes than the total dimension of wavefunction Eqn.~(\ref{HoleWf_N6S7}) which is only $210$.

The unexpected extra zero modes indicates the quasi-hole wavefunction on \cptwo has more than one expression in sharp contrast to the \cpone case. We noted the $\gamma=1$ Jas-Laughlin at $(N,S)=(4,3)$ is an example of zero mode which cannot be written purely as a product of several determinants. We anticipate besides Eqn.~(\ref{HoleWf_N6S7}), general forms of quasi-hole wavefunction should also include those with linear combination of determinants and cancellations, which we decide to discuss more extensively in Section.~\ref{section:math}.

\begin{table*}
\centering
\begin{tabular}{p{1.8cm}|p{1.6cm}|p{1.6cm}|p{1.6cm}|p{1.6cm}|p{1.6cm}|p{1.6cm}|p{1.6cm}|p{1.6cm}|p{1.6cm}}
\hline\hline
 & \hfil I=0 & \hfil I=$\pm$1 & \hfil I=$\pm$2 & \hfil I=$\pm$3 & \hfil I=$\pm$4 & \hfil I=$\pm$5 & \hfil I=$\pm$ 6 & \hfil I=$\pm$7 \\
 \hline
 \hfil Y=~~0 & \hfil 12 (5) & & \hfil 10 (3) & & \hfil 6 (2) & & \hfil 2 &\\
 \hline
 \hfil Y=~~3 & & \hfil 10 (3) & & \hfil 6 (2) & & \hfil 3 & & \\
 \hline
 \hfil Y=$-$3 & & \hfil 10 (3) & & \hfil 8 (2) & & \hfil 3 (1) & & \hfil 1 \\
 \hline
 \hfil Y=~~6 & \hfil 8 (2) & & \hfil 6 (2) & & \hfil 3 & & \hfil 1  & \\
 \hline
 \hfil Y=$-$6 & \hfil 6 (2) & & \hfil 6 (2) & & \hfil 3 (1) & & \hfil 1 (1)  & \\
 \hline
 \hfil Y=~~9 & & \hfil 3 (1) & & \hfil 2 & & & & \\
 \hline
 \hfil Y=$-$9 & & \hfil 3 & & \hfil 2 & & \hfil 1 & & \\
 \hline
 \hfil Y=~~12 & \hfil 1 (1) & & \hfil 1 & & & & & \\
 \hline
 \hfil Y=$-$12 & \hfil 1 & & & & & & & \\
 \hline\hline
\end{tabular}
\caption{Zero mode space dimension of $(N,S)=(6,7)$. \emph{Not all} zero modes can be written as product of three determinants of degree $S=2+2+3$: the number outside the bracket labels the dimension of three determinant product wavefunctions, while that inside the bracket marks the dimension of additional zero modes observed in numerical calculations. The total dimension of three determinant product wavefunctions is $210$, while the total zero mode space dimension is $266$. Empty grids are invalid quantum numbers as they are constrained by Eqn.~(\ref{constrain}).}\label{N6S7}
\end{table*}

\section{$\tSU(3)$ Pseudopotentials\label{section:pp}}
Interactions Eqn.~(\ref{deltaprimeint}) and Eqn.~(\ref{deltaprimeintboson}) are important short ranged interactions. How do we classify generic $\tSU(3)$ symmetric interaction in high dimension? Here, we generalize the Haldane-pseudopotential to higher dimensions, for both two-particle and three-particle interactions. The two-particle pseudopotential was initially derived based on group theoretical analysis by Chyh-Hong Chern {\it et al.} in Ref.~(\onlinecite{CHChen_PRL07}). We use a different approach by using coherent state representations developed in Section.~\ref{section:cp2ll}.

Considering two-particle Hilbert space, as shown in Eqn.~(\ref{2pcoherentwf}), such space is block diagonalized by two-particle coherent states labeled by non-negative integer $J\in[0,S]$. We define the $J-$space projector as $\hat{P}^{(2)}_{S,J}$. The action of any $\tSU(3)$ symmetric interaction can be block-diagonalized into actions within the $J-$subspaces as follows:
\begin{equation}
H = \sum_{J=0}^{S}V^{(2)}_{J}\hat{P}^{(2)}_{S,J},\label{twobodypp}
\end{equation}
where $V^{(2)}_J$ is the interaction decomposition coefficient, which can be defined as the \emph{two-particle $\tSU(3)$ pseudopotential}.

The symmetries are manifest in the Hamiltonian Eqn.~(\ref{twobodypp}). To implement practical calculations, one needs to convert it into the second quantized form such as Eqn.~(\ref{secondquantizedH}). The matrix elements $v_{\bm q\bm q';\bm p\bm p'}$ are straightforwardly derived:
\begin{eqnarray}
    V_{\bm q\bm q';\bm p\bm p'} &=& V^{(2)}_J\langle \bm q\bm q'|\hat{P}^{(2)}_{S,J}|\bm p\bm p'\rangle=V_J^{(2)}\sum_{\bm\alpha}C^{J;\bm\alpha}_{\bm q\bm q'}C^{J;\bm\alpha}_{\bm p\bm p'},\nonumber
\end{eqnarray}
where $C^{J,\bm\alpha}_{\bm p\bm p'}$ is the $\tSU(3)$ Clebsch-Gordan coefficient. The above expression was initially derived in Ref.~(\onlinecite{CHChen_PRL07}).



The three-particle coherent state wavefunction Eqn.~(\ref{threeparticlecoherent}) explicitly shows that the three-particle Hilbert space is block-diagonalized by index $J$ for $\tSU(3)$ symmetric three-particle interaction. Consequently, we have:
\begin{equation}
H = \sum_JV^{(3)}_J\hat{P}^{(3)}_{S,J},\label{3bodypp}
\end{equation}
where $\hat{P}^{(3)}_{J}$ is the three-particle projector that projects three-particle bound states into $J-$subspace. The $V^{(3)}_J$ are defined as the \emph{three-particle $\tSU(3)$ pseudopotential}. And the projector can be similarly represented by the CG coefficients.

\section{Comparison with Mathematical Results \label{section:math}}
It turns out that the problem we are discussing, of fermionic or bosonic wave functions for the Laughlin states in four dimensions, fits nicely into the mathematical framework of commutative algebra of $N$ points in the plane, as studied by M. Haiman \cite{Haiman}.

By ``plane'' here one means a space parameterized by two complex coordinates, so this is the first relation: one can parameterize almost all of $\tcp^2$ by taking general $(u,v)$ in Eq. \ref{triplet}, and solving the constraint to determine $w$.  Thus we can regard
an $N$-particle wavefunction as a function of the $2N$ complex coordinates $u_1,v_1,\ldots,u_N,v_N$.  From Eq. \ref{singleparticlewf} the wavefunctions of interest are polynomial in these variables. Thus we ask:
\begin{question} \label{qv}
Characterize the polynomials in the variables $u_1,v_1,\ldots,u_N,v_N$
which vanish whenever $u_i=u_j$ and $v_i=v_j$ for some $(i,j)$.
\end{question}

The space of such polynomials is an ideal 
$I=\cap_{i<j}(u_i-u_j,v_i-v_j)$ in the polynomial ring $\mathbb C[u_1,v_1,...,u_N,v_N]$ in $2N$ variables.
Now it is easy to see that the analog of the bivariate determinants
Eq.~(\ref{eq:bivandet}) in two variables (equivalently, solving for
the $w_i$'s) are polynomials with this property, but it is not
so obvious that all such polynomials can be obtained this way
(more precisely, are sums of bivariate determinants). Theorem 1.1 of Ref.~(\onlinecite{Haiman}) is precisely this fact, that $I$ coincides with the ideal generated by the bivariate Vandermondes for $N$ points, and indeed the author states that ``this is not an easy theorem.'' We will not even try to explain the proof here, but instead
cite further relevant results from this work.

First, let us compare with the case of complex dimension one.
There the analogous statement was true, namely all polynomials in
$u_1,\ldots,u_N$ which vanish for any $u_i=u_j$
can be obtained as sums of determinants 
$\det_{i,j} u_i^{p_j}$ each multiplied by a polynomial.
But a much simpler statement was also true,
namely all such polynomials can be obtained by multiplying the
Vandermonde determinant $\Delta(u)=\prod_{i<j}(u_i-u_j)$ (the special case with $p_j=j-1$) 
by a single arbitrary polynomial $f$. In other words, the ideal $J=\cap_{i<j}(u_i-u_j)\subseteq\mathbb C[u_1,...,u_N]$ is a principal ideal, i.e. an ideal generated by a single polynomial $\Delta(u)$.
In standard algebraic notations, $J=(\Delta(u))$, where the notation on the right stands for the set of polynomials obtained
by multiplying a given polynomial $\Delta$ by an arbitrary polynomial
$f$. 

To restrict this to totally antisymmetric functions, one need only restrict $f$ to be totally
symmetric.
Physically, this is closely related to exact bose-fermi equivalence
in one dimension -- the bosonic operators (totally symmetric between
particles) act naturally on the free fermion Hilbert space.

Could there be a similar simplification in two variables?
We have translated our question into: is $I$ a principal ideal?
According to Theorem 1.2 of Ref.~(\onlinecite{Haiman}), no:
the situation is more complicated. Fortunately we can broaden our
definitions as follows: let $(\Delta_1,\Delta_2,\ldots,\Delta_k)$ be
the space of polynomials obtained by taking an arbitrary linear
combination $\sum_a f_a \Delta_a$ where $\Delta_a$ are bivariate Vandermondes and the $f_a$'s are general
polynomials.  This is a general ideal, and general results tell us
that this is possible.  

In fact we can be more precise: define the generators of $I$ to be
a basis of elements which cannot be obtained as 
$\sum_a f_a \Delta_a$ where the $\Delta_a\in I$, but the $f_a$ have
no constant part (so, they can be the variables $u_i$, $v_i$ or higher
order polynomials).  Theorem 1.2 tells us that the dimension of this
basis is the $N$'th Catalan number,
\begin{equation}\nonumber
C_N = \frac{1}{N+1}{{2N}\choose{N}} .
\end{equation}

However, the proof is non-constructive, and no explicit choice
for this basis is known except for $N=2,3$.
In physics terms, this tells us that if there is an exact bosonization
in two variables, it will not suffice to let the bosonic operators act on a unique ground state; to get the entire fermionic Hilbert space one will need to start from several (though a finite) number of distinct states.

Let us turn to discuss the FQHE states for the filling fraction $1/\beta$, where $\beta\in\mathbb Z^+$. Here we have two possible definitions -- the physics and the algebraic one. 

{\bf Definition} ({\it algebraic}).  Laughlin states are the polynomials in $\mathbb C[u_1,v_1,...,u_N,v_N]$, which are 

(1) symmetric (for even $\beta$), or anti-symmetric (for odd $\beta$) 
wrt exchanging the coordinates of $N$ points $(u_1,v_1),...,(u_N,v_N)$. 

(2) of partial degree $S$, where partial degree is a sum of top degrees in $u_j$ and $v_j$ (this is independent of the choice of  index $j$ due to (1)).

(3) belonging to $I^\beta$.

The latter condition means that a Laughlin state can be written a linear combination of the form $\sum_{a_!,...,a_\beta} f_{a_1,...,a_\beta}\Delta_{a_1}\cdots\Delta_{a_\beta}$,
where $f_{a_1,...,a_\beta}$ is a (necessarily symmetric) polynomial.

{\bf Definition} ({\it physics}).
 Laughlin states are the polynomials in $\mathbb C[u_1,v_1,...,u_N,v_N]$,
satisfying (1) and (2) above, which are also 

(3') exact ground states of the hamiltonian Eq.\ \eqref{interactH}.

The condition (3') can be reformulated as follows. Let $I^{\langle\beta\rangle}$ denote the ideal of all polynomials in $\mathbb C[u_1,v_1,...,u_N,v_N]$, which belong to $I$ together with all of their partial derivatives of order $\beta-1$ and less, i.e.
\begin{equation}\nonumber
    I^{\langle\beta\rangle}=\bigg\langle f\bigg|\frac{\p^{\bf r}f}{\p u_1^{r_1}...\p v_{2N}^{r_{2N}}}\in I {\rm \;for\; all\;} {\bf r}\in \mathbb N^{2N}, \sum r_n<\beta \bigg\rangle
\end{equation}
then (3') is equivalent to

(3'') Laughlin state is a polynomial in $I^{\langle\beta\rangle}$.

It is easy to see that in complex dimension one the two definitions coincide. 
Luckily, the same property holds in complex dimension two, at least as long as we allow all mixed derivatives of the order up to $\beta-1$ in the hamiltonian Eq.\ \eqref{interactH}. We have  $I^\beta=I^{\langle\beta\rangle}$. The prove is indirect and is a consequence of the property Thm. 1.7. in Ref.~(\onlinecite{Haiman}) of the coincidence of the powers of $I$ with symbolic powers (a notion we do not define here), and the coincidence of symbolic and differential powers (Zariski-Nagata theorem) for the radical ideals \cite{Sullivant}. 

Now we would like to pose
\begin{question}
Compute dimensions of the Hilbert spaces of Laughlin states, as a function of $N,S,\beta$.
\end{question}

In complex dimension one 
we know that all of these states can be obtained by multiplying the
Laughlin state $\Delta^\beta$ by a symmetric function, so it would suffice to count the dimensions of the spaces of symmetric polynomials of given degree. In particular, on a Riemann surface of genus-$g$, we know that there are no Laughlin states for $N>S/\beta+1-g$, at $N=S/\beta+1-g$ their number is $\beta^g$ and the degenaracies for $N<S/\beta+1-g$
have been computed as well\cite{KZ}. 

What is the situation in two variables? Our results show that the situation is rather
complicated, and far from being fully understood.  In some sense,
this question is addressed by Theorem 1.8 and
its Corollary 1.9.  This states that the ideal $I^\beta$, defined
as the product of $\beta$ polynomials each taken from $I$, is just
the $\beta$'th power of the ideal generated by the bivariate determinants.
In other words, every polynomial which vanishes to at least order
$\beta$ when $u_i=u_j$ and $v_i=v_j$ for any pair $(i,j)$, can be
obtained as a sum of terms, each of which is a product of $\beta$ 
bivariate determinants multiplied by some (unconstrained) polynomial.

Now, the discussion in Ref.~(\onlinecite{Haiman}) concerns all
polynomials, with no symmetry or antisymmetry imposed.  Furthermore
the quantum number $S$ is an additional feature of our problem.
It is tempting
to adapt Corollary 1.1 to our physical situation by making the following conjectures:
\begin{conjecture} \label{conj1}
Every symmetric polynomial (for $\beta$ even) or
antisymmetric polynomial (for $\beta$ odd) 
of degree $S$ in each variable,
which vanishes to at least order
$\beta$ when $u_i=u_j$ and $v_i=v_j$ for any pair $(i,j)$, can be
obtained as a sum of terms, each of which is a product of $\beta$ 
bivariate determinants multiplied by a symmetric polynomial.
\end{conjecture}
\begin{conjecture} \label{conj2}
Furthermore each determinant is homogeneous, and the sum of their
degrees $\sum_a S_a = S$.
\end{conjecture}

These conjectures, if true, would give us
a general construction of the FQHE states on $\tcp^2$.  The first step is to list all partitions of the $U(1)$ charge of the form:
\begin{equation}\nonumber
S = \sum_{a=1}^\beta S_a.
\end{equation}

We can then enumerate all of the bivariate determinants of each
required degree $S_a$, of total number 
$(S_a+1)(S_a+2)/2 \choose N$,
using the diagrammatic method.  Finally we combine the choices,
taking into account equivalences which arise if any $S_a=S_b$
for $a\ne b$.

It took us some
time to realize that while Conjecture \ref{conj1} is true
(it follows from the corollary),
Conjecture \ref{conj2} is in fact false.
A simple counterexample is the following: consider the four particle
Jastrow wavefunction $\Psi^J_{\gamma}$ with $\gamma=1$ and thus
vanishing order $\beta=2$; see the second line of Eqn.~(\ref{JasLauWf}). According to the conjecture, it should be
in the ideal generated by products of two bivariate determinants,
schematically
$\sum f_a \Delta \Delta$.
Now since $S=3$, if the determinants are homogeneous, then
one of them must have degree $S_a<2$.
But since there are only three independent states with $S=1$,
all such four particle determinants vanish.

In fact the required sum of products of determinants (which must exist by
Corollary 1.1) is:
\begin{equation} \label{eq:contra}
    u_{123}~u_{124}~u_{134}~u_{234} = 
    \Delta_1^2 - \Delta_2 \Delta_3,
\end{equation}
where the determinants are defined using the following
sets of four indices:
\begin{eqnarray}
\Delta_1 &=& \young(\cdot\cdot,\cdot\cdot)\\
\Delta_2 &=& \young(\cdot,\cdot,\cdot\cdot)\\
\Delta_3 &=& \young(\cdot,\cdot\cdot\cdot)
\end{eqnarray}

Each of these determinants has $S=2$, and the individual products which
appear in Eq. (\ref{eq:contra}) indeed have terms with $S=4$.
However, these terms cancel in the difference, resolving the
contradiction.

We found a similar mismatch at $N=6$ and $S=7$ between the total
count of $\beta=3$ states (266) and the number of states (210) which
can be realized as products of three determinants with $S=2+2+3$.
We believe that the resolution is analogous, that the extra 56
states are sums
of products of three determinants in which the higher degree terms
cancel.

Thus, the construction  of the FQHE states on $\tcp^2$ which we outlined above, does not produce all of the states. To fix this, one would need some understanding of the cancellations we observed, and at the very least a bound on the maximal individual degrees $S_a$'s. 

Finally we note our question about Laughlin states can be posed for any compact K\"ahler manifold, replacing the degree-$S$ polynomials by holomorphic sections of line bundle \cite{Douglas2010} and bivariate Vandermonde determinant by the corresponding Slater determinant \cite{Semyon_RandomNormalMatrices,Semyon_largeNlaughlin,Semyon_highgenus}.

\section{Conclusion and Future Directions \label{section:discussion}}
We have discussed the FQH effect on \cptwo manifold with uniform $\tU(1)$ background magnetic field. We defined two types of Laughlin wavefunction, one of the Determinant type and one of the Vandermonde type. They are respectively shown to be the unique exact zero energy ground states for short ranged two- and three- body interactions, for the $(N,S)$ specified from their wavefunctions such as those listed in Table.~\ref{table_laughlin}. The quasi-hole space degeneracy shows anomalous behavior indicating quasi-hole wavefunction has more than one form, which is different compared to the \cpone usual FQH effects. There are few future research directions that the theory and techniques developed in this work could be useful. For instance a further detailed study of the compressibilities of the two types of Laughlin wavefunctions, as well as their low-energy excitations \cite{gmpl} are interesting. In contrast to the two-dimension, high dimension may support membrane-like excitations \cite{BERNEVIG2002185,4dqh_membrane}. It may also be possible to extend this work to other quantum Hall states including the paired states \cite{MoreReadState} and gapless states \cite{HalperinLeeRead,Son,Jie_MonteCarlo,scottjiehaldane,Jie_Dirac,SongyangPuCFL}. A thorough mathematical understanding of quasi-hole wavefunctions besides Section.~\ref{section:math} is an open questions. Last but not least, searching for experimental realizations of interacting physics in high dimensional Landau levels \cite{4dqh_coldatom,Zilberberg:2018aa,Lohse:2018aa} are important future directions.

\section*{Acknowledgments}
We are grateful to Vladimir Dotsenko, Benoit Estienne and Nicolas Regnault for helpful discussions. We acknowledge the open-source diagonalization code \emph{Diagham}, especially the \cptwo FQH part developed by Cecile Repellin and Nicolas Regnault, for inspirations. Numerical studies presented in this paper were based on the codes developed by the authors involved in this work. The work of S.K. was partly supported by the IdEx program and the USIAS Fellowship of the University of Strasbourg, and the ANR-20-CE40-0017 grant. The Flatiron Institute is a division of the Simons Foundation.

\bibliography{fqh.bib}

\begin{thebibliography}{51}%
\makeatletter
\providecommand \@ifxundefined [1]{%
 \@ifx{#1\undefined}
}%
\providecommand \@ifnum [1]{%
 \ifnum #1\expandafter \@firstoftwo
 \else \expandafter \@secondoftwo
 \fi
}%
\providecommand \@ifx [1]{%
 \ifx #1\expandafter \@firstoftwo
 \else \expandafter \@secondoftwo
 \fi
}%
\providecommand \natexlab [1]{#1}%
\providecommand \enquote  [1]{``#1''}%
\providecommand \bibnamefont  [1]{#1}%
\providecommand \bibfnamefont [1]{#1}%
\providecommand \citenamefont [1]{#1}%
\providecommand \href@noop [0]{\@secondoftwo}%
\providecommand \href [0]{\begingroup \@sanitize@url \@href}%
\providecommand \@href[1]{\@@startlink{#1}\@@href}%
\providecommand \@@href[1]{\endgroup#1\@@endlink}%
\providecommand \@sanitize@url [0]{\catcode `\\12\catcode `\$12\catcode
  `\&12\catcode `\#12\catcode `\^12\catcode `\_12\catcode `\%12\relax}%
\providecommand \@@startlink[1]{}%
\providecommand \@@endlink[0]{}%
\providecommand \url  [0]{\begingroup\@sanitize@url \@url }%
\providecommand \@url [1]{\endgroup\@href {#1}{\urlprefix }}%
\providecommand \urlprefix  [0]{URL }%
\providecommand \Eprint [0]{\href }%
\providecommand \doibase [0]{http://dx.doi.org/}%
\providecommand \selectlanguage [0]{\@gobble}%
\providecommand \bibinfo  [0]{\@secondoftwo}%
\providecommand \bibfield  [0]{\@secondoftwo}%
\providecommand \translation [1]{[#1]}%
\providecommand \BibitemOpen [0]{}%
\providecommand \bibitemStop [0]{}%
\providecommand \bibitemNoStop [0]{.\EOS\space}%
\providecommand \EOS [0]{\spacefactor3000\relax}%
\providecommand \BibitemShut  [1]{\csname bibitem#1\endcsname}%
\let\auto@bib@innerbib\@empty
\bibitem [{\citenamefont {Zhang}\ and\ \citenamefont {Hu}(2001)}]{HuZhang01}%
  \BibitemOpen
  \bibfield  {author} {\bibinfo {author} {\bibfnamefont {S.-C.}\ \bibnamefont
  {Zhang}}\ and\ \bibinfo {author} {\bibfnamefont {J.}~\bibnamefont {Hu}},\
  }\href {\doibase 10.1126/science.294.5543.823} {\bibfield  {journal}
  {\bibinfo  {journal} {Science}\ }\textbf {\bibinfo {volume} {294}},\ \bibinfo
  {pages} {823} (\bibinfo {year} {2001})},\ \Eprint
  {http://arxiv.org/abs/https://science.sciencemag.org/content/294/5543/823.full.pdf}
  {https://science.sciencemag.org/content/294/5543/823.full.pdf} \BibitemShut
  {NoStop}%
\bibitem [{\citenamefont {Karabali}\ and\ \citenamefont
  {Nair}(2002)}]{KARABALI02}%
  \BibitemOpen
  \bibfield  {author} {\bibinfo {author} {\bibfnamefont {D.}~\bibnamefont
  {Karabali}}\ and\ \bibinfo {author} {\bibfnamefont {V.}~\bibnamefont
  {Nair}},\ }\href {\doibase https://doi.org/10.1016/S0550-3213(02)00634-X}
  {\bibfield  {journal} {\bibinfo  {journal} {Nuclear Physics B}\ }\textbf
  {\bibinfo {volume} {641}},\ \bibinfo {pages} {533} (\bibinfo {year}
  {2002})}\BibitemShut {NoStop}%
\bibitem [{\citenamefont {Stormer}\ \emph {et~al.}(1999)\citenamefont
  {Stormer}, \citenamefont {Tsui},\ and\ \citenamefont
  {Gossard}}]{RevModPhys.71.S298}%
  \BibitemOpen
  \bibfield  {author} {\bibinfo {author} {\bibfnamefont {H.~L.}\ \bibnamefont
  {Stormer}}, \bibinfo {author} {\bibfnamefont {D.~C.}\ \bibnamefont {Tsui}}, \
  and\ \bibinfo {author} {\bibfnamefont {A.~C.}\ \bibnamefont {Gossard}},\
  }\href {\doibase 10.1103/RevModPhys.71.S298} {\bibfield  {journal} {\bibinfo
  {journal} {Rev. Mod. Phys.}\ }\textbf {\bibinfo {volume} {71}},\ \bibinfo
  {pages} {S298} (\bibinfo {year} {1999})}\BibitemShut {NoStop}%
\bibitem [{\citenamefont {Hasan}\ and\ \citenamefont
  {Kane}(2010)}]{RevModPhys.82.3045}%
  \BibitemOpen
  \bibfield  {author} {\bibinfo {author} {\bibfnamefont {M.~Z.}\ \bibnamefont
  {Hasan}}\ and\ \bibinfo {author} {\bibfnamefont {C.~L.}\ \bibnamefont
  {Kane}},\ }\href {\doibase 10.1103/RevModPhys.82.3045} {\bibfield  {journal}
  {\bibinfo  {journal} {Rev. Mod. Phys.}\ }\textbf {\bibinfo {volume} {82}},\
  \bibinfo {pages} {3045} (\bibinfo {year} {2010})}\BibitemShut {NoStop}%
\bibitem [{\citenamefont {Laughlin}(1983)}]{laughlin}%
  \BibitemOpen
  \bibfield  {author} {\bibinfo {author} {\bibfnamefont {R.~B.}\ \bibnamefont
  {Laughlin}},\ }\href {\doibase 10.1103/PhysRevLett.50.1395} {\bibfield
  {journal} {\bibinfo  {journal} {Phys. Rev. Lett.}\ }\textbf {\bibinfo
  {volume} {50}},\ \bibinfo {pages} {1395} (\bibinfo {year}
  {1983})}\BibitemShut {NoStop}%
\bibitem [{\citenamefont {Nayak}\ \emph {et~al.}(2008)\citenamefont {Nayak},
  \citenamefont {Simon}, \citenamefont {Stern}, \citenamefont {Freedman},\ and\
  \citenamefont {Das~Sarma}}]{RevModPhys.80.1083}%
  \BibitemOpen
  \bibfield  {author} {\bibinfo {author} {\bibfnamefont {C.}~\bibnamefont
  {Nayak}}, \bibinfo {author} {\bibfnamefont {S.~H.}\ \bibnamefont {Simon}},
  \bibinfo {author} {\bibfnamefont {A.}~\bibnamefont {Stern}}, \bibinfo
  {author} {\bibfnamefont {M.}~\bibnamefont {Freedman}}, \ and\ \bibinfo
  {author} {\bibfnamefont {S.}~\bibnamefont {Das~Sarma}},\ }\href {\doibase
  10.1103/RevModPhys.80.1083} {\bibfield  {journal} {\bibinfo  {journal} {Rev.
  Mod. Phys.}\ }\textbf {\bibinfo {volume} {80}},\ \bibinfo {pages} {1083}
  (\bibinfo {year} {2008})}\BibitemShut {NoStop}%
\bibitem [{\citenamefont {Karabali}\ and\ \citenamefont
  {Nair}(2006)}]{Karabali_2006}%
  \BibitemOpen
  \bibfield  {author} {\bibinfo {author} {\bibfnamefont {D.}~\bibnamefont
  {Karabali}}\ and\ \bibinfo {author} {\bibfnamefont {V.~P.}\ \bibnamefont
  {Nair}},\ }\href {\doibase 10.1088/0305-4470/39/41/s05} {\bibfield  {journal}
  {\bibinfo  {journal} {Journal of Physics A: Mathematical and General}\
  }\textbf {\bibinfo {volume} {39}},\ \bibinfo {pages} {12735} (\bibinfo {year}
  {2006})}\BibitemShut {NoStop}%
\bibitem [{\citenamefont {Karabali}\ and\ \citenamefont
  {Nair}(2016)}]{Karabali16}%
  \BibitemOpen
  \bibfield  {author} {\bibinfo {author} {\bibfnamefont {D.}~\bibnamefont
  {Karabali}}\ and\ \bibinfo {author} {\bibfnamefont {V.~P.}\ \bibnamefont
  {Nair}},\ }\href {\doibase 10.1103/PhysRevD.94.024022} {\bibfield  {journal}
  {\bibinfo  {journal} {Phys. Rev. D}\ }\textbf {\bibinfo {volume} {94}},\
  \bibinfo {pages} {024022} (\bibinfo {year} {2016})}\BibitemShut {NoStop}%
\bibitem [{\citenamefont {Karabali}(2020)}]{Karabali20}%
  \BibitemOpen
  \bibfield  {author} {\bibinfo {author} {\bibfnamefont {D.}~\bibnamefont
  {Karabali}},\ }\href {\doibase 10.1103/PhysRevD.102.025016} {\bibfield
  {journal} {\bibinfo  {journal} {Phys. Rev. D}\ }\textbf {\bibinfo {volume}
  {102}},\ \bibinfo {pages} {025016} (\bibinfo {year} {2020})}\BibitemShut
  {NoStop}%
\bibitem [{\citenamefont {Douglas}\ and\ \citenamefont
  {Klevtsov}(2010)}]{Douglas2010}%
  \BibitemOpen
  \bibfield  {author} {\bibinfo {author} {\bibfnamefont {M.~R.}\ \bibnamefont
  {Douglas}}\ and\ \bibinfo {author} {\bibfnamefont {S.}~\bibnamefont
  {Klevtsov}},\ }\href {\doibase 10.1007/s00220-009-0915-0} {\bibfield
  {journal} {\bibinfo  {journal} {Communications in Mathematical Physics}\
  }\textbf {\bibinfo {volume} {293}},\ \bibinfo {pages} {205} (\bibinfo {year}
  {2010})}\BibitemShut {NoStop}%
\bibitem [{\citenamefont {Estienne}\ \emph {et~al.}(2021)\citenamefont
  {Estienne}, \citenamefont {Oblak},\ and\ \citenamefont
  {Stéphan}}]{4dqhedge_ergodic}%
  \BibitemOpen
  \bibfield  {author} {\bibinfo {author} {\bibfnamefont {B.}~\bibnamefont
  {Estienne}}, \bibinfo {author} {\bibfnamefont {B.}~\bibnamefont {Oblak}}, \
  and\ \bibinfo {author} {\bibfnamefont {J.-M.}\ \bibnamefont {Stéphan}},\
  }\href {\doibase 10.21468/SciPostPhys.11.1.016} {\bibfield  {journal}
  {\bibinfo  {journal} {SciPost Phys.}\ }\textbf {\bibinfo {volume} {11}},\
  \bibinfo {pages} {16} (\bibinfo {year} {2021})}\BibitemShut {NoStop}%
\bibitem [{\citenamefont {Casteill}\ and\ \citenamefont
  {Nersessian}(2003)}]{4dqh_particlecp3}%
  \BibitemOpen
  \bibfield  {author} {\bibinfo {author} {\bibfnamefont {P.-Y.}\ \bibnamefont
  {Casteill}}\ and\ \bibinfo {author} {\bibfnamefont {A.}~\bibnamefont
  {Nersessian}},\ }\href {\doibase 10.1016/j.physletb.2003.09.008} {\bibfield
  {journal} {\bibinfo  {journal} {Physics Letters B}\ }\textbf {\bibinfo
  {volume} {574}},\ \bibinfo {pages} {121} (\bibinfo {year}
  {2003})}\BibitemShut {NoStop}%
\bibitem [{\citenamefont {Ozawa}\ and\ \citenamefont
  {Price}(2019)}]{review_syndim}%
  \BibitemOpen
  \bibfield  {author} {\bibinfo {author} {\bibfnamefont {T.}~\bibnamefont
  {Ozawa}}\ and\ \bibinfo {author} {\bibfnamefont {H.~M.}\ \bibnamefont
  {Price}},\ }\href {\doibase 10.1038/s42254-019-0045-3} {\bibfield  {journal}
  {\bibinfo  {journal} {Nature Reviews Physics}\ }\textbf {\bibinfo {volume}
  {1}},\ \bibinfo {pages} {349} (\bibinfo {year} {2019})}\BibitemShut {NoStop}%
\bibitem [{\citenamefont {Price}\ \emph {et~al.}(2015)\citenamefont {Price},
  \citenamefont {Zilberberg}, \citenamefont {Ozawa}, \citenamefont
  {Carusotto},\ and\ \citenamefont {Goldman}}]{4dqh_coldatom}%
  \BibitemOpen
  \bibfield  {author} {\bibinfo {author} {\bibfnamefont {H.~M.}\ \bibnamefont
  {Price}}, \bibinfo {author} {\bibfnamefont {O.}~\bibnamefont {Zilberberg}},
  \bibinfo {author} {\bibfnamefont {T.}~\bibnamefont {Ozawa}}, \bibinfo
  {author} {\bibfnamefont {I.}~\bibnamefont {Carusotto}}, \ and\ \bibinfo
  {author} {\bibfnamefont {N.}~\bibnamefont {Goldman}},\ }\href {\doibase
  10.1103/PhysRevLett.115.195303} {\bibfield  {journal} {\bibinfo  {journal}
  {Phys. Rev. Lett.}\ }\textbf {\bibinfo {volume} {115}},\ \bibinfo {pages}
  {195303} (\bibinfo {year} {2015})}\BibitemShut {NoStop}%
\bibitem [{\citenamefont {Zilberberg}\ \emph {et~al.}(2018)\citenamefont
  {Zilberberg}, \citenamefont {Huang}, \citenamefont {Guglielmon},
  \citenamefont {Wang}, \citenamefont {Chen}, \citenamefont {Kraus},\ and\
  \citenamefont {Rechtsman}}]{Zilberberg:2018aa}%
  \BibitemOpen
  \bibfield  {author} {\bibinfo {author} {\bibfnamefont {O.}~\bibnamefont
  {Zilberberg}}, \bibinfo {author} {\bibfnamefont {S.}~\bibnamefont {Huang}},
  \bibinfo {author} {\bibfnamefont {J.}~\bibnamefont {Guglielmon}}, \bibinfo
  {author} {\bibfnamefont {M.}~\bibnamefont {Wang}}, \bibinfo {author}
  {\bibfnamefont {K.~P.}\ \bibnamefont {Chen}}, \bibinfo {author}
  {\bibfnamefont {Y.~E.}\ \bibnamefont {Kraus}}, \ and\ \bibinfo {author}
  {\bibfnamefont {M.~C.}\ \bibnamefont {Rechtsman}},\ }\href {\doibase
  10.1038/nature25011} {\bibfield  {journal} {\bibinfo  {journal} {Nature}\
  }\textbf {\bibinfo {volume} {553}},\ \bibinfo {pages} {59} (\bibinfo {year}
  {2018})}\BibitemShut {NoStop}%
\bibitem [{\citenamefont {Lohse}\ \emph {et~al.}(2018)\citenamefont {Lohse},
  \citenamefont {Schweizer}, \citenamefont {Price}, \citenamefont
  {Zilberberg},\ and\ \citenamefont {Bloch}}]{Lohse:2018aa}%
  \BibitemOpen
  \bibfield  {author} {\bibinfo {author} {\bibfnamefont {M.}~\bibnamefont
  {Lohse}}, \bibinfo {author} {\bibfnamefont {C.}~\bibnamefont {Schweizer}},
  \bibinfo {author} {\bibfnamefont {H.~M.}\ \bibnamefont {Price}}, \bibinfo
  {author} {\bibfnamefont {O.}~\bibnamefont {Zilberberg}}, \ and\ \bibinfo
  {author} {\bibfnamefont {I.}~\bibnamefont {Bloch}},\ }\href {\doibase
  10.1038/nature25000} {\bibfield  {journal} {\bibinfo  {journal} {Nature}\
  }\textbf {\bibinfo {volume} {553}},\ \bibinfo {pages} {55} (\bibinfo {year}
  {2018})}\BibitemShut {NoStop}%
\bibitem [{\citenamefont {Kraus}\ \emph {et~al.}(2013)\citenamefont {Kraus},
  \citenamefont {Ringel},\ and\ \citenamefont
  {Zilberberg}}]{4dqh_quasicrystal}%
  \BibitemOpen
  \bibfield  {author} {\bibinfo {author} {\bibfnamefont {Y.~E.}\ \bibnamefont
  {Kraus}}, \bibinfo {author} {\bibfnamefont {Z.}~\bibnamefont {Ringel}}, \
  and\ \bibinfo {author} {\bibfnamefont {O.}~\bibnamefont {Zilberberg}},\
  }\href {\doibase 10.1103/PhysRevLett.111.226401} {\bibfield  {journal}
  {\bibinfo  {journal} {Phys. Rev. Lett.}\ }\textbf {\bibinfo {volume} {111}},\
  \bibinfo {pages} {226401} (\bibinfo {year} {2013})}\BibitemShut {NoStop}%
\bibitem [{\citenamefont {Haldane}(1983)}]{haldanehierarchy}%
  \BibitemOpen
  \bibfield  {author} {\bibinfo {author} {\bibfnamefont {F.~D.~M.}\
  \bibnamefont {Haldane}},\ }\href {\doibase 10.1103/PhysRevLett.51.605}
  {\bibfield  {journal} {\bibinfo  {journal} {Phys. Rev. Lett.}\ }\textbf
  {\bibinfo {volume} {51}},\ \bibinfo {pages} {605} (\bibinfo {year}
  {1983})}\BibitemShut {NoStop}%
\bibitem [{\citenamefont {Greiter}(2011)}]{GreiterCP1}%
  \BibitemOpen
  \bibfield  {author} {\bibinfo {author} {\bibfnamefont {M.}~\bibnamefont
  {Greiter}},\ }\href {\doibase 10.1103/PhysRevB.83.115129} {\bibfield
  {journal} {\bibinfo  {journal} {Phys. Rev. B}\ }\textbf {\bibinfo {volume}
  {83}},\ \bibinfo {pages} {115129} (\bibinfo {year} {2011})}\BibitemShut
  {NoStop}%
\bibitem [{\citenamefont {Chern}\ and\ \citenamefont
  {Lee}(2007)}]{CHChen_PRL07}%
  \BibitemOpen
  \bibfield  {author} {\bibinfo {author} {\bibfnamefont {C.-H.}\ \bibnamefont
  {Chern}}\ and\ \bibinfo {author} {\bibfnamefont {D.-H.}\ \bibnamefont
  {Lee}},\ }\href {\doibase 10.1103/PhysRevLett.98.066804} {\bibfield
  {journal} {\bibinfo  {journal} {Phys. Rev. Lett.}\ }\textbf {\bibinfo
  {volume} {98}},\ \bibinfo {pages} {066804} (\bibinfo {year}
  {2007})}\BibitemShut {NoStop}%
\bibitem [{\citenamefont {Chern}(2007)}]{CHChen_AOP07}%
  \BibitemOpen
  \bibfield  {author} {\bibinfo {author} {\bibfnamefont {C.-H.}\ \bibnamefont
  {Chern}},\ }\href {\doibase https://doi.org/10.1016/j.aop.2006.11.011}
  {\bibfield  {journal} {\bibinfo  {journal} {Annals of Physics}\ }\textbf
  {\bibinfo {volume} {322}},\ \bibinfo {pages} {2485} (\bibinfo {year}
  {2007})}\BibitemShut {NoStop}%
\bibitem [{\citenamefont {Chern}(2010)}]{CHChen_PRB10}%
  \BibitemOpen
  \bibfield  {author} {\bibinfo {author} {\bibfnamefont {C.-H.}\ \bibnamefont
  {Chern}},\ }\href {\doibase 10.1103/PhysRevB.81.115123} {\bibfield  {journal}
  {\bibinfo  {journal} {Phys. Rev. B}\ }\textbf {\bibinfo {volume} {81}},\
  \bibinfo {pages} {115123} (\bibinfo {year} {2010})}\BibitemShut {NoStop}%
\bibitem [{\citenamefont {Kitaev}(2003)}]{KITAEV20032}%
  \BibitemOpen
  \bibfield  {author} {\bibinfo {author} {\bibfnamefont {A.}~\bibnamefont
  {Kitaev}},\ }\href {\doibase https://doi.org/10.1016/S0003-4916(02)00018-0}
  {\bibfield  {journal} {\bibinfo  {journal} {Annals of Physics}\ }\textbf
  {\bibinfo {volume} {303}},\ \bibinfo {pages} {2} (\bibinfo {year}
  {2003})}\BibitemShut {NoStop}%
\bibitem [{\citenamefont {Haiman}\ and\ \citenamefont {Miller}(2004)}]{Haiman}%
  \BibitemOpen
  \bibfield  {author} {\bibinfo {author} {\bibfnamefont {M.}~\bibnamefont
  {Haiman}}\ and\ \bibinfo {author} {\bibfnamefont {E.}~\bibnamefont
  {Miller}},\ }\enquote {\bibinfo {title} {Commutative algebra of n points in
  the plane},}\ in\ \href {\doibase 10.1017/CBO9780511756382.006} {\emph
  {\bibinfo {booktitle} {Trends in Commutative Algebra}}},\ \bibinfo {series
  and number} {Mathematical Sciences Research Institute Publications},\
  \bibinfo {editor} {edited by\ \bibinfo {editor} {\bibfnamefont {L.~L.}\
  \bibnamefont {Avramov}}, \bibinfo {editor} {\bibfnamefont {M.}~\bibnamefont
  {Green}}, \bibinfo {editor} {\bibfnamefont {C.}~\bibnamefont {Huneke}},
  \bibinfo {editor} {\bibfnamefont {K.~E.}\ \bibnamefont {Smith}}, \ and\
  \bibinfo {editor} {\bibfnamefont {B.}~\bibnamefont {Sturmfels}}}\ (\bibinfo
  {publisher} {Cambridge University Press},\ \bibinfo {year} {2004})\ p.\
  \bibinfo {pages} {153–180}\BibitemShut {NoStop}%
\bibitem [{\citenamefont {Wen}(2017)}]{XGW_review}%
  \BibitemOpen
  \bibfield  {author} {\bibinfo {author} {\bibfnamefont {X.-G.}\ \bibnamefont
  {Wen}},\ }\href {\doibase 10.1103/RevModPhys.89.041004} {\bibfield  {journal}
  {\bibinfo  {journal} {Rev. Mod. Phys.}\ }\textbf {\bibinfo {volume} {89}},\
  \bibinfo {pages} {041004} (\bibinfo {year} {2017})}\BibitemShut {NoStop}%
\bibitem [{\citenamefont {Janowitz}\ \emph {et~al.}(2013)\citenamefont
  {Janowitz}, \citenamefont {Yanagisawa}, \citenamefont {Eisaki}, \citenamefont
  {Trallero-Giner},\ and\ \citenamefont {Wen}}]{Janowitz:2013aa}%
  \BibitemOpen
  \bibfield  {author} {\bibinfo {author} {\bibfnamefont {C.}~\bibnamefont
  {Janowitz}}, \bibinfo {author} {\bibfnamefont {T.}~\bibnamefont
  {Yanagisawa}}, \bibinfo {author} {\bibfnamefont {H.}~\bibnamefont {Eisaki}},
  \bibinfo {author} {\bibfnamefont {C.}~\bibnamefont {Trallero-Giner}}, \ and\
  \bibinfo {author} {\bibfnamefont {X.-G.}\ \bibnamefont {Wen}},\ }\href
  {\doibase 10.1155/2013/198710} {\bibfield  {journal} {\bibinfo  {journal}
  {ISRN Condensed Matter Physics}\ }\textbf {\bibinfo {volume} {2013}},\
  \bibinfo {pages} {198710} (\bibinfo {year} {2013})}\BibitemShut {NoStop}%
\bibitem [{\citenamefont {{Zhang}}\ and\ \citenamefont
  {{Ye}}(2021)}]{PengYe2104}%
  \BibitemOpen
  \bibfield  {author} {\bibinfo {author} {\bibfnamefont {Z.-F.}\ \bibnamefont
  {{Zhang}}}\ and\ \bibinfo {author} {\bibfnamefont {P.}~\bibnamefont {{Ye}}},\
  }\href@noop {} {\bibfield  {journal} {\bibinfo  {journal} {arXiv e-prints}\
  ,\ \bibinfo {eid} {arXiv:2104.07067}} (\bibinfo {year} {2021})},\ \Eprint
  {http://arxiv.org/abs/2104.07067} {arXiv:2104.07067 [hep-th]} \BibitemShut
  {NoStop}%
\bibitem [{\citenamefont {WEN}(1990)}]{Wen_TO_rigidstates}%
  \BibitemOpen
  \bibfield  {author} {\bibinfo {author} {\bibfnamefont {X.~G.}\ \bibnamefont
  {WEN}},\ }\href {\doibase 10.1142/S0217979290000139} {\bibfield  {journal}
  {\bibinfo  {journal} {International Journal of Modern Physics B}\ }\textbf
  {\bibinfo {volume} {04}},\ \bibinfo {pages} {239} (\bibinfo {year} {1990})},\
  \Eprint {http://arxiv.org/abs/https://doi.org/10.1142/S0217979290000139}
  {https://doi.org/10.1142/S0217979290000139} \BibitemShut {NoStop}%
\bibitem [{\citenamefont {Wen}(1995)}]{Wen_TO_edgeFQH}%
  \BibitemOpen
  \bibfield  {author} {\bibinfo {author} {\bibfnamefont {X.-G.}\ \bibnamefont
  {Wen}},\ }\href {\doibase 10.1080/00018739500101566} {\bibfield  {journal}
  {\bibinfo  {journal} {Advances in Physics}\ }\textbf {\bibinfo {volume}
  {44}},\ \bibinfo {pages} {405} (\bibinfo {year} {1995})},\ \Eprint
  {http://arxiv.org/abs/https://doi.org/10.1080/00018739500101566}
  {https://doi.org/10.1080/00018739500101566} \BibitemShut {NoStop}%
\bibitem [{\citenamefont {{Xiao-Gang Wen}}\ \emph {et~al.}(1994)\citenamefont
  {{Xiao-Gang Wen}}, \citenamefont {{Yong-Shi Wu}},\ and\ \citenamefont
  {{Yasuhiro Hatsugai}}}]{Wen_Int}%
  \BibitemOpen
  \bibfield  {author} {\bibinfo {author} {\bibnamefont {{Xiao-Gang Wen}}},
  \bibinfo {author} {\bibnamefont {{Yong-Shi Wu}}}, \ and\ \bibinfo {author}
  {\bibnamefont {{Yasuhiro Hatsugai}}},\ }\href {\doibase
  https://doi.org/10.1016/0550-3213(94)90442-1} {\bibfield  {journal} {\bibinfo
   {journal} {Nuclear Physics B}\ }\textbf {\bibinfo {volume} {422}},\ \bibinfo
  {pages} {476} (\bibinfo {year} {1994})}\BibitemShut {NoStop}%
\bibitem [{\citenamefont {Bernevig}\ \emph {et~al.}(2002)\citenamefont
  {Bernevig}, \citenamefont {Chern}, \citenamefont {Hu}, \citenamefont
  {Toumbas},\ and\ \citenamefont {Zhang}}]{BERNEVIG2002185}%
  \BibitemOpen
  \bibfield  {author} {\bibinfo {author} {\bibfnamefont {B.~A.}\ \bibnamefont
  {Bernevig}}, \bibinfo {author} {\bibfnamefont {C.-H.}\ \bibnamefont {Chern}},
  \bibinfo {author} {\bibfnamefont {J.-P.}\ \bibnamefont {Hu}}, \bibinfo
  {author} {\bibfnamefont {N.}~\bibnamefont {Toumbas}}, \ and\ \bibinfo
  {author} {\bibfnamefont {S.-C.}\ \bibnamefont {Zhang}},\ }\href {\doibase
  https://doi.org/10.1006/aphy.2002.6292} {\bibfield  {journal} {\bibinfo
  {journal} {Annals of Physics}\ }\textbf {\bibinfo {volume} {300}},\ \bibinfo
  {pages} {185} (\bibinfo {year} {2002})}\BibitemShut {NoStop}%
\bibitem [{\citenamefont {Heckman}\ and\ \citenamefont
  {Tizzano}(2018)}]{4dqh_membrane}%
  \BibitemOpen
  \bibfield  {author} {\bibinfo {author} {\bibfnamefont {J.~J.}\ \bibnamefont
  {Heckman}}\ and\ \bibinfo {author} {\bibfnamefont {L.}~\bibnamefont
  {Tizzano}},\ }\href {\doibase 10.1007/JHEP05(2018)120} {\bibfield  {journal}
  {\bibinfo  {journal} {Journal of High Energy Physics}\ }\textbf {\bibinfo
  {volume} {2018}},\ \bibinfo {pages} {120} (\bibinfo {year}
  {2018})}\BibitemShut {NoStop}%
\bibitem [{\citenamefont {Zhang}\ \emph {et~al.}(1990)\citenamefont {Zhang},
  \citenamefont {Feng},\ and\ \citenamefont {Gilmore}}]{RMP_coherentstate}%
  \BibitemOpen
  \bibfield  {author} {\bibinfo {author} {\bibfnamefont {W.-M.}\ \bibnamefont
  {Zhang}}, \bibinfo {author} {\bibfnamefont {D.~H.}\ \bibnamefont {Feng}}, \
  and\ \bibinfo {author} {\bibfnamefont {R.}~\bibnamefont {Gilmore}},\ }\href
  {\doibase 10.1103/RevModPhys.62.867} {\bibfield  {journal} {\bibinfo
  {journal} {Rev. Mod. Phys.}\ }\textbf {\bibinfo {volume} {62}},\ \bibinfo
  {pages} {867} (\bibinfo {year} {1990})}\BibitemShut {NoStop}%
\bibitem [{\citenamefont {Luis}(2008)}]{Luis_2008}%
  \BibitemOpen
  \bibfield  {author} {\bibinfo {author} {\bibfnamefont {A.}~\bibnamefont
  {Luis}},\ }\href {\doibase 10.1088/1751-8113/41/49/495302} {\bibfield
  {journal} {\bibinfo  {journal} {Journal of Physics A: Mathematical and
  Theoretical}\ }\textbf {\bibinfo {volume} {41}},\ \bibinfo {pages} {495302}
  (\bibinfo {year} {2008})}\BibitemShut {NoStop}%
\bibitem [{\citenamefont {Luis}(2005{\natexlab{a}})}]{Luis05PRA1}%
  \BibitemOpen
  \bibfield  {author} {\bibinfo {author} {\bibfnamefont {A.}~\bibnamefont
  {Luis}},\ }\href {\doibase 10.1103/PhysRevA.71.063815} {\bibfield  {journal}
  {\bibinfo  {journal} {Phys. Rev. A}\ }\textbf {\bibinfo {volume} {71}},\
  \bibinfo {pages} {063815} (\bibinfo {year} {2005}{\natexlab{a}})}\BibitemShut
  {NoStop}%
\bibitem [{\citenamefont {Luis}(2005{\natexlab{b}})}]{Luis05PRA2}%
  \BibitemOpen
  \bibfield  {author} {\bibinfo {author} {\bibfnamefont {A.}~\bibnamefont
  {Luis}},\ }\href {\doibase 10.1103/PhysRevA.71.023810} {\bibfield  {journal}
  {\bibinfo  {journal} {Phys. Rev. A}\ }\textbf {\bibinfo {volume} {71}},\
  \bibinfo {pages} {023810} (\bibinfo {year} {2005}{\natexlab{b}})}\BibitemShut
  {NoStop}%
\bibitem [{\citenamefont {Trugman}\ and\ \citenamefont
  {Kivelson}(1985)}]{Trugman_Kivelson}%
  \BibitemOpen
  \bibfield  {author} {\bibinfo {author} {\bibfnamefont {S.~A.}\ \bibnamefont
  {Trugman}}\ and\ \bibinfo {author} {\bibfnamefont {S.}~\bibnamefont
  {Kivelson}},\ }\href {\doibase 10.1103/PhysRevB.31.5280} {\bibfield
  {journal} {\bibinfo  {journal} {Phys. Rev. B}\ }\textbf {\bibinfo {volume}
  {31}},\ \bibinfo {pages} {5280} (\bibinfo {year} {1985})}\BibitemShut
  {NoStop}%
\bibitem [{\citenamefont {Girvin}\ \emph {et~al.}(1985)\citenamefont {Girvin},
  \citenamefont {MacDonald},\ and\ \citenamefont {Platzman}}]{gmpl}%
  \BibitemOpen
  \bibfield  {author} {\bibinfo {author} {\bibfnamefont {S.~M.}\ \bibnamefont
  {Girvin}}, \bibinfo {author} {\bibfnamefont {A.~H.}\ \bibnamefont
  {MacDonald}}, \ and\ \bibinfo {author} {\bibfnamefont {P.~M.}\ \bibnamefont
  {Platzman}},\ }\href {\doibase 10.1103/PhysRevLett.54.581} {\bibfield
  {journal} {\bibinfo  {journal} {Phys. Rev. Lett.}\ }\textbf {\bibinfo
  {volume} {54}},\ \bibinfo {pages} {581} (\bibinfo {year} {1985})}\BibitemShut
  {NoStop}%
\bibitem [{\citenamefont {Girvin}\ \emph {et~al.}(1986)\citenamefont {Girvin},
  \citenamefont {MacDonald},\ and\ \citenamefont {Platzman}}]{gmpb}%
  \BibitemOpen
  \bibfield  {author} {\bibinfo {author} {\bibfnamefont {S.~M.}\ \bibnamefont
  {Girvin}}, \bibinfo {author} {\bibfnamefont {A.~H.}\ \bibnamefont
  {MacDonald}}, \ and\ \bibinfo {author} {\bibfnamefont {P.~M.}\ \bibnamefont
  {Platzman}},\ }\href {\doibase 10.1103/PhysRevB.33.2481} {\bibfield
  {journal} {\bibinfo  {journal} {Phys. Rev. B}\ }\textbf {\bibinfo {volume}
  {33}},\ \bibinfo {pages} {2481} (\bibinfo {year} {1986})}\BibitemShut
  {NoStop}%
\bibitem [{\citenamefont {Sullivant}(2008)}]{Sullivant}%
  \BibitemOpen
  \bibfield  {author} {\bibinfo {author} {\bibfnamefont {S.}~\bibnamefont
  {Sullivant}},\ }\href {\doibase 10.1016/j.jalgebra.2007.09.024} {\bibfield
  {journal} {\bibinfo  {journal} {Journal of Algebra}\ }\textbf {\bibinfo
  {volume} {319}},\ \bibinfo {pages} {115} (\bibinfo {year}
  {2008})}\BibitemShut {NoStop}%
\bibitem [{\citenamefont {Klevtsov}\ and\ \citenamefont {Zvonkine}(2021)}]{KZ}%
  \BibitemOpen
  \bibfield  {author} {\bibinfo {author} {\bibfnamefont {S.}~\bibnamefont
  {Klevtsov}}\ and\ \bibinfo {author} {\bibfnamefont {D.}~\bibnamefont
  {Zvonkine}},\ }\href@noop {} {} (\bibinfo {year} {2021}),\ \Eprint
  {http://arxiv.org/abs/2105.00989} {arXiv:2105.00989 [cond-mat.str-el]}
  \BibitemShut {NoStop}%
\bibitem [{\citenamefont {Klevtsov}(2014)}]{Semyon_RandomNormalMatrices}%
  \BibitemOpen
  \bibfield  {author} {\bibinfo {author} {\bibfnamefont {S.}~\bibnamefont
  {Klevtsov}},\ }\href {\doibase 10.1007/JHEP01(2014)133} {\bibfield  {journal}
  {\bibinfo  {journal} {Journal of High Energy Physics}\ }\textbf {\bibinfo
  {volume} {2014}},\ \bibinfo {pages} {133} (\bibinfo {year}
  {2014})}\BibitemShut {NoStop}%
\bibitem [{\citenamefont {Klevtsov}(2016)}]{Semyon_largeNlaughlin}%
  \BibitemOpen
  \bibfield  {author} {\bibinfo {author} {\bibfnamefont {S.}~\bibnamefont
  {Klevtsov}},\ }\href@noop {} {\bibfield  {journal} {\bibinfo  {journal}
  {Trav. Math.}\ }\textbf {\bibinfo {volume} {24}},\ \bibinfo {pages} {63}
  (\bibinfo {year} {2016})},\ \Eprint {http://arxiv.org/abs/1608.02928}
  {arXiv:1608.02928 [cond-mat.str-el]} \BibitemShut {NoStop}%
\bibitem [{\citenamefont {Klevtsov}(2019)}]{Semyon_highgenus}%
  \BibitemOpen
  \bibfield  {author} {\bibinfo {author} {\bibfnamefont {S.}~\bibnamefont
  {Klevtsov}},\ }\href {\doibase 10.1007/s00220-019-03318-6} {\bibfield
  {journal} {\bibinfo  {journal} {Communications in Mathematical Physics}\
  }\textbf {\bibinfo {volume} {367}},\ \bibinfo {pages} {837} (\bibinfo {year}
  {2019})}\BibitemShut {NoStop}%
\bibitem [{\citenamefont {Moore}\ and\ \citenamefont
  {Read}(1991)}]{MoreReadState}%
  \BibitemOpen
  \bibfield  {author} {\bibinfo {author} {\bibfnamefont {G.}~\bibnamefont
  {Moore}}\ and\ \bibinfo {author} {\bibfnamefont {N.}~\bibnamefont {Read}},\
  }\href {\doibase https://doi.org/10.1016/0550-3213(91)90407-O} {\bibfield
  {journal} {\bibinfo  {journal} {Nuclear Physics B}\ }\textbf {\bibinfo
  {volume} {360}},\ \bibinfo {pages} {362} (\bibinfo {year}
  {1991})}\BibitemShut {NoStop}%
\bibitem [{\citenamefont {Halperin}\ \emph {et~al.}(1993)\citenamefont
  {Halperin}, \citenamefont {Lee},\ and\ \citenamefont
  {Read}}]{HalperinLeeRead}%
  \BibitemOpen
  \bibfield  {author} {\bibinfo {author} {\bibfnamefont {B.~I.}\ \bibnamefont
  {Halperin}}, \bibinfo {author} {\bibfnamefont {P.~A.}\ \bibnamefont {Lee}}, \
  and\ \bibinfo {author} {\bibfnamefont {N.}~\bibnamefont {Read}},\ }\href
  {\doibase 10.1103/PhysRevB.47.7312} {\bibfield  {journal} {\bibinfo
  {journal} {Phys. Rev. B}\ }\textbf {\bibinfo {volume} {47}},\ \bibinfo
  {pages} {7312} (\bibinfo {year} {1993})}\BibitemShut {NoStop}%
\bibitem [{\citenamefont {Son}(2015)}]{Son}%
  \BibitemOpen
  \bibfield  {author} {\bibinfo {author} {\bibfnamefont {D.~T.}\ \bibnamefont
  {Son}},\ }\href {\doibase 10.1103/PhysRevX.5.031027} {\bibfield  {journal}
  {\bibinfo  {journal} {Phys. Rev. X}\ }\textbf {\bibinfo {volume} {5}},\
  \bibinfo {pages} {031027} (\bibinfo {year} {2015})}\BibitemShut {NoStop}%
\bibitem [{\citenamefont {Wang}\ \emph {et~al.}(2019)\citenamefont {Wang},
  \citenamefont {Geraedts}, \citenamefont {Rezayi},\ and\ \citenamefont
  {Haldane}}]{Jie_MonteCarlo}%
  \BibitemOpen
  \bibfield  {author} {\bibinfo {author} {\bibfnamefont {J.}~\bibnamefont
  {Wang}}, \bibinfo {author} {\bibfnamefont {S.~D.}\ \bibnamefont {Geraedts}},
  \bibinfo {author} {\bibfnamefont {E.~H.}\ \bibnamefont {Rezayi}}, \ and\
  \bibinfo {author} {\bibfnamefont {F.~D.~M.}\ \bibnamefont {Haldane}},\ }\href
  {\doibase 10.1103/PhysRevB.99.125123} {\bibfield  {journal} {\bibinfo
  {journal} {Phys. Rev. B}\ }\textbf {\bibinfo {volume} {99}},\ \bibinfo
  {pages} {125123} (\bibinfo {year} {2019})}\BibitemShut {NoStop}%
\bibitem [{\citenamefont {Geraedts}\ \emph {et~al.}(2018)\citenamefont
  {Geraedts}, \citenamefont {Wang}, \citenamefont {Rezayi},\ and\ \citenamefont
  {Haldane}}]{scottjiehaldane}%
  \BibitemOpen
  \bibfield  {author} {\bibinfo {author} {\bibfnamefont {S.~D.}\ \bibnamefont
  {Geraedts}}, \bibinfo {author} {\bibfnamefont {J.}~\bibnamefont {Wang}},
  \bibinfo {author} {\bibfnamefont {E.~H.}\ \bibnamefont {Rezayi}}, \ and\
  \bibinfo {author} {\bibfnamefont {F.~D.~M.}\ \bibnamefont {Haldane}},\ }\href
  {\doibase 10.1103/PhysRevLett.121.147202} {\bibfield  {journal} {\bibinfo
  {journal} {Phys. Rev. Lett.}\ }\textbf {\bibinfo {volume} {121}},\ \bibinfo
  {pages} {147202} (\bibinfo {year} {2018})}\BibitemShut {NoStop}%
\bibitem [{\citenamefont {Wang}(2019)}]{Jie_Dirac}%
  \BibitemOpen
  \bibfield  {author} {\bibinfo {author} {\bibfnamefont {J.}~\bibnamefont
  {Wang}},\ }\href {\doibase 10.1103/PhysRevLett.122.257203} {\bibfield
  {journal} {\bibinfo  {journal} {Phys. Rev. Lett.}\ }\textbf {\bibinfo
  {volume} {122}},\ \bibinfo {pages} {257203} (\bibinfo {year}
  {2019})}\BibitemShut {NoStop}%
\bibitem [{\citenamefont {Pu}\ \emph {et~al.}(2018)\citenamefont {Pu},
  \citenamefont {Fremling},\ and\ \citenamefont {Jain}}]{SongyangPuCFL}%
  \BibitemOpen
  \bibfield  {author} {\bibinfo {author} {\bibfnamefont {S.}~\bibnamefont
  {Pu}}, \bibinfo {author} {\bibfnamefont {M.}~\bibnamefont {Fremling}}, \ and\
  \bibinfo {author} {\bibfnamefont {J.~K.}\ \bibnamefont {Jain}},\ }\href
  {\doibase 10.1103/PhysRevB.98.075304} {\bibfield  {journal} {\bibinfo
  {journal} {Phys. Rev. B}\ }\textbf {\bibinfo {volume} {98}},\ \bibinfo
  {pages} {075304} (\bibinfo {year} {2018})}\BibitemShut {NoStop}%
\end{thebibliography}%
\end{document}